\documentclass[showpacs,superscriptaddress,preprintnumbers,amsmath,
amssymb,nofootinbib]{revtex4}
\usepackage{amsmath,amssymb,latexsym}
\usepackage{graphicx,bm}
\usepackage{dcolumn}    
\usepackage{epsf,psfig}
\begin{document}
%
%
%
%
\newcommand{\sinc}{ {\mathrm{sinc}} } 
\newcommand{\hf}{ {\hat{f}} }
\newcommand{\Rdots}{\rotatebox[origin=c]{-45}{\vdots}}
%
%
%
%
%
\title{ Probing anisotropies of gravitational-wave backgrounds
  \\
  with a space-based interferometer III:
  \\
  \it{ Reconstruction of a high-frequency skymap  } }

\author{Atsushi Taruya} \email{ataruya_at_utap.phys.s.u-tokyo.ac.jp}
\affiliation{ Research Center for the Early Universe~(RESCEU), School
  of Science, The University of Tokyo, Tokyo 113-0033, Japan }

\preprint{UTAP-559, RESCEU-17/06}
\begin{abstract}
  We develop a numerical scheme to make a 
  high-frequency skymap of gravitational-wave backgrounds (GWBs) 
  observed via space-based 
  interferometer. Based on the cross-correlation technique, the 
  intensity distribution of anisotropic GWB can be directly reconstructed 
  from the time-ordered data of cross-correlation signals, with full 
  knowledge of detector's antenna pattern functions. 
  We demonstrate how the planned space interferometer, LISA, can make a 
  skymap of GWB for a specific example of anisotropic signals. 
  At the frequency higher than the characteristic frequency 
  $f_*=1/(2\pi\,L)$, where $L$ is the arm-length of the 
  detector, the reconstructed skymap free from the 
  instrumental noise potentially reaches the angular resolution  
  up to the multipoles $\ell\sim10$. The presence of instrumental 
  noises degrades the angular resolution. The resultant skymap has 
  angular resolution with multipoles $\ell\leq 6\sim7$ for the 
  anisotropic signals with signal-to-noise ratio S/N$>5$. 
\end{abstract}

\maketitle

\section{Introduction}
\label{sec:intro}

The incoherent superposition of gravitational waves coming from 
many unresolved sources and/or diffuse-sources may form a 
stochastic background of gravitational waves. While such a tiny signal  
is in nature random, their statistical 
properties contain valuable cosmological information about the 
astrophysical phenomena and the cosmic expansion history. 
Potentially, the extremely early stage of the universe can be 
directly probed by using the gravitational-wave background (GWB), 
beyond the last scattering surface of cosmological microwave background.

Aiming at detecting the tiny fluctuations,  
several future missions of space-based interferometers have been proposed. 
Among these, Laser Interferometer Space Antenna (LISA) will be a 
first-generation space-based detector launched in the next decade 
\footnote{URL, http://lisa.jpl.nasa.gov/,~ http://sci.esa.int/science-e/www/area/index.cfm?fareaid=27 }. 
The main target of LISA is the low-frequency gravitational waves 
of astrophysical origin around $1\sim10$mHz. It is expected that 
large population of Galactic binaries produces a strong signal of GWB at 
$f\lesssim 1$mHz \cite{Hils:1990hg,Bender:1997bc,Nelemans:2001hp}. 
Further, the extragalactic signals of GWB would be 
detected at high-frequency band \cite{Schneider:2000sg,Farmer:2003pa}.  
On the other hand, second-generation space interferometers, 
the DECI-hertz interferometer Gravitational wave Observatory (DECIGO) 
\cite{Seto:2001qf} and the Big-Bang Observer 
(BBO)\footnote{URL, http://universe.nasa.gov/program/bbo.html} 
\cite{Phinney:2003}, have been proposed as follow-on mission of 
LISA. The main target of these missions is the primordial GWB 
produced during the inflationary epoch 
(e.g., \cite{Turner:1996ck,Boyle:2005ug,Smith:2005mm,Chongchitnan:2006pe}). 
For this purpose, the 
observational window around $0.1\sim1$Hz is considered to be suitable 
for direct detection. With future technology accessible to the next 
two decades, direct detection will be possible for the inflationary 
GWB with amplitude $\Omega_{\rm gw}\sim10^{-16}$ 
\cite{Ungarelli:2005qb,Seto:2005qy,Kudoh:2005as}.

The detection of GWB in the next few decades will open a 
new subject of cosmology and thereby the infrastructure such 
as a new data analysis technique is needed to be exploited. 
Among various interesting problems, making a skymap of GWB 
is a key piece in observational cosmology. 
In our previous study, we have investigated the directional sensitivity 
of space interferometer to the anisotropy of GWB 
(Ref.\cite{Kudoh:2004he}, hereafter paper I). 
It turned out that the angular sensitivity of gravitational-wave detector to 
the anisotropic GWBs sensitively depends on the geometric 
configuration of space interferometer as well as the signal processing. 
Further, based on the correlation analysis, a method for direct 
reconstruction of 
GWB skymap has been exploited (Ref.\cite{Taruya:2005yf}, hereafter paper II). 
Employing the perturbative technique, a low-frequency skymap of GWB can be 
directly reconstructed (for the reconstruction in the low-frequency limit, 
see Ref.\cite{Seto:2004np}).

The aim of the present paper is to extend our previous study to 
the direct reconstruction of a high-frequency skymap beyond the 
low-frequency approximation. We give a simple reconstruction method,  
in which the problem is reduced to find a solution of the linear 
algebraic system. While the governing linear equation often becomes 
under-determined, with a full numerical treatment, the present scheme 
gives an approximate but reliable solution for skymap reconstruction. 
As a demonstration, we consider a specific example for 
the reconstruction problem using the space interferometer LISA. 
We show that at the high-frequency band higher than the characteristic 
frequency $f_*=1/(2\pi\,L)\sim9.52$mHz, the LISA can make  
a skymap of GWB with angular resolution up to the multipoles $\ell\sim10$. 
The presence of instrumental noises significantly degrades 
the angular resolution and thereby the reconstructed result of 
the multipole coefficients for GWB skymap includes a larger error.  
However, with the signal-to-noise  ratio S/N$>5$, the resultant skymap  
has angular resolution with multipoles $\ell\leq 6\sim7$.

The paper is organized as follows. In Sec.~\ref{sec:method}, 
based on the cross-correlation technique to detect the stochastic signals, 
we briefly review how to extract the information of 
anisotropies of the GWBs and present the simple reconstruction 
scheme, which is capable of applying to the high-frequency skymap. 
Several distinctions between the high-frequency and the 
low-frequency cases are discussed. In Sec.~\ref{sec:demo}, the 
reconstruction scheme is demonstrated in the specific model of 
the anisotropic GWB, i.e., Galactic white dwarf binaries. The quality 
of the reconstructed skymap is quantified 
in both the noise-free and the noisy cases. 
Finally, Sec.~\ref{sec:conclusion} is devoted to the summary and 
the discussion.

\section{Methodology}
\label{sec:method}

\subsection{Map-making problem}
\label{subsec:map-making}

Before addressing the reconstruction scheme, 
we briefly summarize how one can detect the anisotropies of GWB 
via space interferometer. 

The planned space interferometer, LISA and also the next generation 
detectors DECIGO/BBO constitute several spacecrafts, each of which 
exchanges laser beams with the others. Combining these laser pulses, 
it is possible to synthesize the various output streams which are 
sensitive (or insensitive) to the gravitational-wave signal. 
In particular, technique to synthesize data streams 
canceling the laser frequency noise is known as time-delay 
interferometry (TDI),  which is crucial for LISA mission 
(\cite{Armstrong:1999, Tinto:2003vj, Shaddock:2003dj}, see 
\cite{Tinto:2004wu} for a review).  
In the present paper, signals produced by TDI technique are used to 
demonstrate the skymap reconstruction.

The multiple data streams constructed from specific combinations 
of laser pulses can be used to detect the GWBs through 
the correlation analysis. Suppose that one obtains the two output 
streams denoted by $s_I(t)$ and $s_J(t)$, the correlation analysis 
can be made by calculating the ensemble average 
$C_{IJ}\equiv\langle s_I(t)s_J(t)\rangle$. Naively, one may think that 
the correlation signal $C_{IJ}$ becomes time-independent 
if both the signal and the instrumental noises obeys 
stationary random process. However, this is not entirely correct, because 
the sensitivity of gravitational-wave detector has specific angular 
response to the GWB. Further, the motion of space interferometer is inherently 
non-stationary. For LISA, the constellation of three space crafts orbits 
around the Sun with a period of one sidereal year 
and the orientation of the gravitational-wave detector 
gradually changes relative to the sky. Hence, in presence of the 
anisotropic GWB, the amplitude of correlation signal cannot be constant, 
but varies in time. This is key ingredient for reconstruction of the 
GWB skymap. The basic equation characterizing the non-stationarity of 
the correlation signal can be written in the Fourier domain as 
(paper I, II, \cite{Giampieri:1997,Allen:1997gp,Cornish:2001hg,Ungarelli:2001xu,Seto:2004ji}): 
\begin{equation}
C_{IJ}(f,t)=\int\, \frac{d\mathbf{\Omega}}{4\pi}\,\,
S_{h}(f,\mathbf{\Omega})\,\,\mathcal{F}_{IJ}(f,\mathbf{\Omega};t),   
\label{eq:corr_A}
\end{equation}
where the correlation signal $C(f,t)$ is related with 
the one in the time-domain through the expression, 
$C_{IJ}=\int df C_{IJ}(f,t)$. The quantity $S_h$ 
represents the power spectral density of GWB, which is, in general,  
the unknown function of the frequency and the sky position. On the other 
hand,  the function $\mathcal{F}_{IJ}$ is called the antenna 
pattern function, which characterizes the angular response of 
gravitational-wave detector, and the precise functional form of it is 
known from the configuration and the orientation of the detector 
(see Appendix \ref{appendix:antenna_pattern}).

Eq.~(\ref{eq:corr_A}) implies that the luminosity distribution of GWB 
 $S_h(f,\mathbf{\Omega})$ is reconstructed by deconvolving the all-sky 
integral 
of antenna pattern function from the time-series data $C_{IJ}(f,t)$. 
To see this more explicitly, we decompose the antenna pattern function and 
the luminosity distribution of GWB into spherical harmonics in a 
sky-fixed coordinate system:  
\begin{eqnarray}
    S_h(|f|,\, \mathbf{\Omega}) 
    =  \sum_{\ell, m} \,\,[p_{\ell m} (f)]^* \,\,
    Y_{\ell m}^* ( \mathbf{\Omega}), 
\quad
    \mathcal{F}_{IJ}(f,\, \mathbf{\Omega};\,t)
    = 
    \sum_{\ell, m} \,\, a_{\ell m} (f,t) \,\,
    Y_{\ell m} ( \mathbf{\Omega}). 
    \label{eq:Y_lm expansion of S and F}
\end{eqnarray}
Note that the properties of spherical harmonics yield 
$p_{\ell m}^*=(-1)^m p_{\ell,-m}$ and 
$a_{\ell m}^*=(-1)^{\ell-m}a_{\ell,-m}$, where the latter property comes from 
${\mathcal{F}}_{IJ}^* (f,  \Omega;t) = {\mathcal{F}}_{IJ} (f, - \Omega;t)$ 
(paper I). Substituting (\ref{eq:Y_lm expansion of S and F}) into 
(\ref{eq:corr_A}) becomes
\begin{eqnarray}
     C_{IJ}(t,f) = \frac{1}{4\pi}\, \sum_{\ell m}\,\, a_{\ell m}(f,\,t)
     \,\,\left[ p_{\ell m}(f) \right]^*.   
 \label{eq:corr_B}
 \end{eqnarray}
Thus, the problem to reconstruct a GWB skymap reduces to the linear 
algebraic problem. That is, collecting the correlation signals measured at 
different times and solving the couples of linear equations, 
the multipole coefficient of GWB $p_{\ell m}(f)$ can be obtained 
under a full knowledge of time-dependent coefficients $a_{\ell m}(f,t)$.

Several important remarks should be mentioned in order. First, 
the accessible multipole coefficients $p_{\ell m}$ are severely 
restricted by the angular sensitivity of antenna pattern functions. 
According to paper I, the space interferometer LISA is typically 
sensitive to the multipole coefficients $\ell\lesssim5$ of 
the anisotropic GWBs in the  low-frequency regime 
and to the multipoles $\ell\lesssim10$ in the high-frequency regime. Second, 
the above linear system is generally either over-constrained or 
under-determined. In the presence of the instrumental 
noises, this deconvolution problem tends to become under-determined system 
(paper II). As a consequence, exact reconstruction is no longer than 
possible in practice and we need to exploit an approximate method to 
reconstruct the GWB skymap with a limited angular resolution.

\subsection{Reconstruction scheme}
\label{subsec:reconstruction}

In what follows, assuming a prior knowledge of the 
time-dependent antenna pattern functions, 
we present simple reconstruction method for a GWB skymap from 
time-modulated correlation signals. In paper II, employing the perturbative 
expansion of the antenna pattern functions, reconstruction method for  
a low-frequency skymap has been presented. In the present paper, we 
especially focus on the high-frequency skymap. Here, the high-frequency 
means that the wavelength of the gravitational waves 
is comparable or shorter than the arm-length of the detector 
(or separation between the two detectors that produce the correlation 
signals), where the low-frequency approximation of the antenna pattern 
breaks down.

The basic strategy to make a high-frequency skymap is almost the same as in 
paper II. Suppose that for a given frequency $f$, one obtains 
the discrete time-series data for correlation signals  
as $C_{IJ}(f,t_i)$ $(i=1,2,\cdots,N)$. 
We then write Eq.~(\ref{eq:corr_B}) in the matrix form as:  
\begin{eqnarray}
\mathbf{c}(f)  = \mathbf{A}(f)  \cdot \mathbf{p}(f),     
\label{eq:c=Ap}
\end{eqnarray}
where the vector ${\mathbf {c}}$ has $N$ columns, each of which contain 
the correlation signal $C_{IJ}(f;t_i)$. 
The vector $\mathbf{p}(f)$ represents the unknown multipole coefficients 
of the GWB spectrum and in each column, we have $p_{\ell m}^*$. Thus, 
if one truncates the spherical harmonic expansion with the multipole 
$\ell_{\rm max}$, the total number of the elements in vector 
$\mathbf{p}(f)$ becomes $(\ell_{\rm max}+1)^2$.  
On the other hand, the matrix ${\mathbf {A}}$ contains 
the multipole coefficients of antenna pattern functions and  
the matrix has $a_{\ell m}(f,t_i)$ in each element. With 
the truncation multipole $\ell_{\rm max}$, 
the quantity ${\mathbf {A}}$ forms a $(\ell_{\rm max}+1)^2\times N$ matrix.

As mentioned in previous subsection, 
the linear system (\ref{eq:c=Ap}) tends to become an 
under-determined system, i.e., $(\ell_{\rm max}+1)^2>N$, and 
$\mathbf{A}$ is generally a rectangular matrix. In such a situation, 
unique and exact solution of the linear equations (\ref{eq:c=Ap}) 
cannot be obtained. 
In paper II, approximate treatment to solve the equations (\ref{eq:c=Ap}) 
has been presented based on the idea of least-squares method. In linear 
system, this approximation is expressed in the following form:
\begin{eqnarray}
{\mathbf {p}}_{\rm approx}(f)= {\mathbf {A}}^{+}(f)
{\cdot \mathbf c}(f), 
\label{eq:approx}
\end{eqnarray}
where ${\mathbf {A}}^{+}$ is the pseudo-inverse matrix of Moore-Penrose 
type. The explicit expression is determined from the singular-value 
decomposition of the matrix ${\mathbf {A}}$, i.e., 
${\mathbf A} =U^{\dagger}\cdot \mbox{diag}[w_i]\cdot V$. Then we have 
(e.g., \cite{Press:NRf77})
\begin{eqnarray}
\mathbf{A}^{+} = V^{\dagger}\cdot \mbox{diag}[w_i^{-1}]\cdot U. 
\label{eq:pseudo-inverse}
\end{eqnarray}
In paper II, owing to the perturbative expansion, 
the multipole coefficients of antenna pattern function can be computed 
analytically up to the multipole $\ell=5$
and the matrix $\mathbf{A}$ is constructed in an analytic manner. 
It is shown that the least-squares solution (\ref{eq:approx}) 
provides an accurate approximation for multipole moments $p_{\ell m}$ 
in both overdetermined and under-determined cases.

In the high-frequency regime which we are interested in, 
the angular resolution of antenna pattern function can be improved and 
one expects that detectable multipole coefficients increase 
compared to those of the low-frequency skymap. 
As a price, it is no longer possible to 
compute the multipole coefficients $a_{\ell m}$ analytically and the 
the spherical harmonic expansion of antenna pattern function 
becomes a fully numerical task. In the present paper, we use the 
Fortran package, SPHEREPACK 3.1 \cite{Adams:2003}, 
to compute the time-dependent 
coefficients $a_{\ell m}$. With a full numerical treatment 
to evaluate Eq.~(\ref{eq:approx}), the reliability of the methodology 
will be tested in a simple reconstruction problem (Sec.~\ref{sec:demo}).

In principle, the methodology presented in paper II can work well 
even in the map-making problem of the 
high-frequency GWBs. Nonetheless, to further reduce the numerical error, 
the least-squares method (\ref{eq:approx}) may be applied  
by imposing the reality condition of the GWB intensity map, i.e., 
$p_{\ell m}(f)=(-1)^m p_{\ell,-m}^*(f)$. To do this, 
we introduce real quantities $q_{\ell m}$ and $r_{\ell m}$ 
and divide the multipole coefficient $p_{\ell m}$ 
into the real and the imaginary parts as 
\begin{eqnarray}
p_{\ell m}&=&q_{\ell m}+ i\,r_{\ell m}, 
\nonumber\\
p_{\ell,-m}&=&(-1)^m\,\left\{ q_{\ell m} - i\,r_{\ell m}\right\}
\nonumber
\end{eqnarray}
for $m>0$ and 
\begin{eqnarray}
p_{\ell,0}&=&q_{\ell 0}.
\nonumber 
\end{eqnarray}
Then, Eq.(\ref{eq:corr_B}) is rewritten with
\begin{eqnarray}
 C_{IJ}(f;t) &=& \frac{1}{4\pi}\, \sum_{\ell=0}\,\, 
    \left[ a_{\ell 0}\,\, q_{\ell 0} + \sum_{m=1}^{\ell} 
        \left\{ a_{\ell m}+(-1)^m\,a_{\ell,-m} \right\} q_{\ell m}
    -i\,\sum_{m=1}^{\ell} 
    \left\{ a_{\ell m}-(-1)^m\,a_{\ell,-m} \right\} r_{\ell m}
\right] 
\nonumber \\
&=& \frac{1}{4\pi}\,\sum_{\ell=0}\,\vec{a}_{\ell}(f;t)\cdot\vec{p}_{\ell}(f),
\end{eqnarray}
where the vectors $\vec{a}_{\ell}$ and $\vec{p}_{\ell}$ have the 
$(2\ell+1)$ columns, whose explicit expressions are 
\begin{eqnarray}
\vec{a}_{\ell}=\left(
\begin{array}{l}
a_{\ell 0} \\
\vdots \\
a_{\ell m}+(-1)^m\,a_{\ell,-m} \\
\vdots \\
\vdots \\
-i\left\{a_{\ell m}-(-1)^m\,a_{\ell,-m}\right\}\\
\vdots \\
\end{array}
\right), 
\quad\quad
\vec{p}_{\ell}=\left(
\begin{array}{c}
q_{\ell 0} \\
\vdots\\
q_{\ell m} \\
\vdots\\
\vdots\\
r_{\ell m}\\
\vdots\\
\end{array}
\right),
\end{eqnarray}
where $m$ runs from $1$ to $\ell$. Hence, truncating the summation of 
$\ell$ by $\ell_{\rm max}$ and dividing the time-ordered signals into 
the $N$ subsections, the matrix $\mathbf{A}$ and the vector $\mathbf{p}$ 
can be written explicitly
\begin{eqnarray}
\mathbf{A}=\frac{1}{4\pi}\,
\left(
\begin{array}{ccccc}
\vec{a}_0(t_1)&\cdots&\vec{a}_{\ell}(t_1)&
\cdots&\vec{a}_{\ell_{\rm max}}(t_1)\\
    \vdots     &  \ddots & \vdots & \ddots & \vdots  \\
\vec{a}_0(t_N) &  \cdots & \vec{a}_{\ell}(t_N) & 
\cdots&\vec{a}_{\ell_{\rm max}}(t_N)\\
\end{array}
\right), 
\quad\quad
\mathbf{p}=\left(
\begin{array}{c}
\vec{p}_0 \\
\vdots    \\
\vec{p}_{\ell_{\rm max}} 
\end{array}
\right).
\label{eq:new_matrix_vector}
\end{eqnarray}
For a reliable numerical calculation, 
we further divide the matrix $\mathbf{A}$ in Eq.~(\ref{eq:new_matrix_vector}) 
into the real and the imaginary parts and 
apply the least-squares approximation (\ref{eq:approx}) 
to the linear system (\ref{eq:c=Ap}).

\section{Demonstration}
\label{sec:demo}

\begin{figure}[t]
\begin{center}
\includegraphics[width=10cm,clip,angle=0]{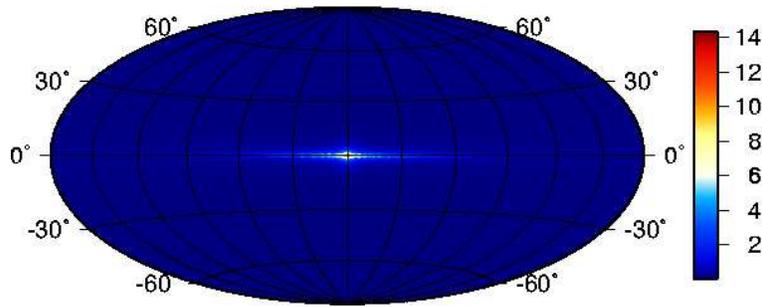}
\end{center}

\vspace*{-0.3cm}

    \caption{Full resolution skymap for the simple model of Galactic GWB 
      (paper II, \cite{Seto:2004ji}). 
      The intensity distribution $S_h(f,\mathbf{\Omega})=P(\mathbf{\Omega})$ 
      is depicted in the Galactic coordinate system. Here, the all-sky 
      integral of intensity distribution is normalized to unity, i.e., 
      $\int d\mathbf{\Omega}\,P(\mathbf{\Omega})=1$. }
    \label{fig:full_skymap}
\end{figure}

In this section, general reconstruction method presented in previous 
section is demonstrated in the case of LISA detector and 
the map-making capability is examined for a specific example 
of anisotropic GWB source.

It is expected that the low-frequency band of LISA have 
a strong anisotropic signals of GWB by the Galactic population of 
unresolved binaries. On the other hand, GWBs of high-frequency regime 
could be dominated by the extragalactic origin, which mainly comes from 
the cosmological population of white dwarf binaries 
\cite{Schneider:2000sg,Farmer:2003pa}. 
While the signal produced by them would reach the detectable level of LISA 
sensitivity, it is uncertain whether the strong 
anisotropic component exists or not. In this paper, just for illustrative 
purpose, we consider the Galactic GWB as a high-frequency source of GWB 
and try to reconstruct the intensity distribution of GWB.

Fig.~\ref{fig:full_skymap} shows the projected skymap of Galactic GWB. 
Here, we use the simple model of Galactic GWB described in paper II 
(see also \cite{Seto:2004ji}), in 
which we assumed that the intensity distribution of GWB 
just traces the Galactic stellar 
distribution observed via infrared photometry \cite{Binney:1996sv}\footnote{To be precise, the Galactic stellar distribution is modeled by the fitting function given in Ref.\cite{Binney:1996sv}, which consists of the triaxial bulge and the disk components.}. The 
intensity distribution shown in Fig.~\ref{fig:full_skymap} has 
disk-like structure with a strong peak at Galactic center. 
Spherical harmonic analysis of GWB skymap reveals that 
the dominant contribution to the intensity distribution comes from the 
lower multipoles with $\ell\lesssim4$, but the contribution from 
the higher multipoles still remains significant. Even at the multipole 
$\ell\sim15$, it possesses the $10\%$ power relative to the monoopole 
component (see Table III and Fig.~12 in paper II). 
In this sense, it is a good exercise to diagnose the map-making 
capability of present reconstruction scheme.

As we mentioned in Sec.~\ref{subsec:map-making}, signal processing 
technique of LISA is called the TDI, which produces the various output 
signals canceling the laser frequency noise 
by combining the time-delayed six laser pulses. Among these, optimal 
TDI signals referred to as the $A$, $E$ and $T$ variables are especially 
suitable for the detection of GWB via the correlation analysis, 
because these outputs are basically free from the noise correlation 
\cite{Prince:2002hp,Nayak:2002ir}. Here, we use the optimal TDIs 
to produce the correlation signals. The response functions for these 
variables are presented in the equal armlength case in 
Appendix \ref{appendix:antenna_pattern}. 
For the reconstruction of a high-frequency
skymap, the cross-correlation signals $AE$, $AT$ and $ET$ can be the most 
sensitive signals of GWB and the reconstructed skymap 
improves the angular resolution up to the multipoles 
$\ell\sim 8-10$ around the frequencies $f\sim 2-10f_*$, where 
the characteristic frequency of LISA is given by 
$f_*=1/(2\pi\, L)\simeq9.52$mHz.

In what follows, specifically focusing on the frequency 
$f=3f_*\simeq28.6$mHz, we present the reconstruction results 
separately in idealistically noise-free case (Sec. \ref{subsec:noise-free}) 
and in realistic case taking account of the instrumental noises 
(Sec. \ref{subsec:noisy}).

\subsection{Noise-free case}
\label{subsec:noise-free}

\begin{figure}[t]
\begin{center}
\includegraphics[width=10cm,clip,angle=0]{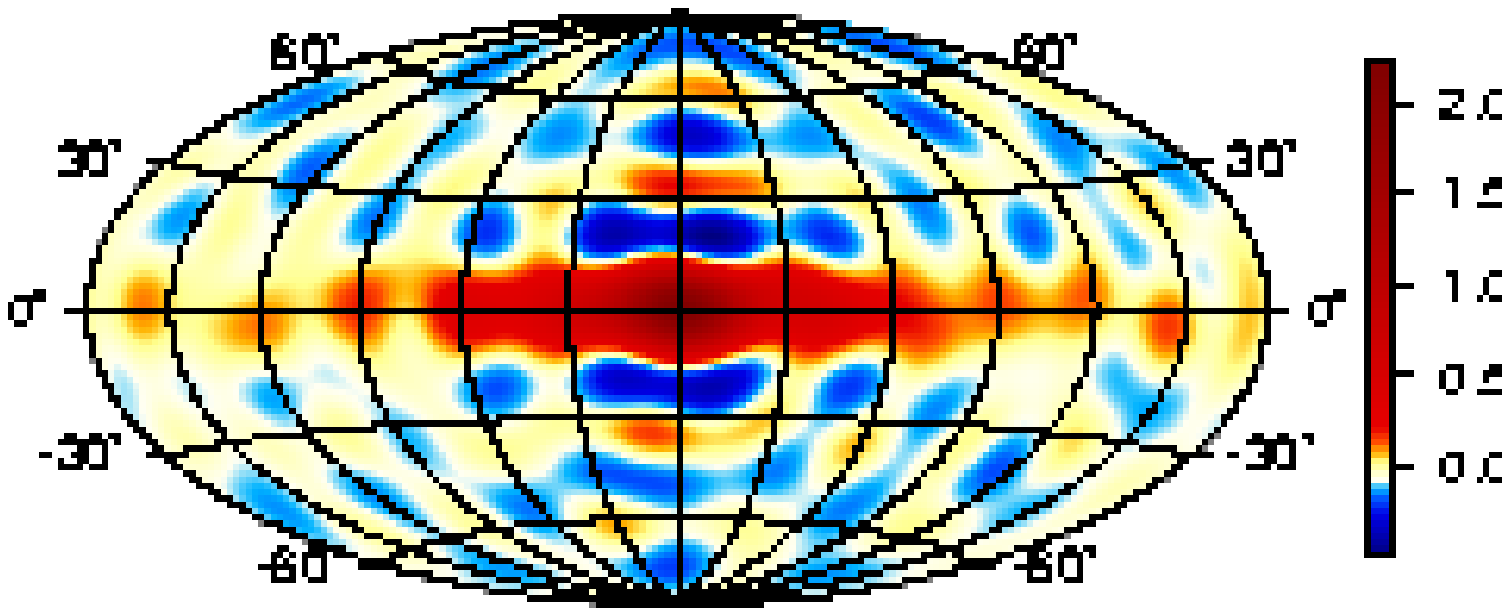}
\end{center}

\vspace*{-0.3cm}

    \caption{Reconstructed skymap at the frequency $f=3f_*\simeq28.6$mHz 
      in the absence of instrumental noises. 
      The cutoff parameter of singular values $w_{\rm cut}$ 
      is set to $10^{-7}$. }
    \label{fig:skymap_noisefree1}
\begin{center}
\includegraphics[width=5.8cm,clip,angle=0]{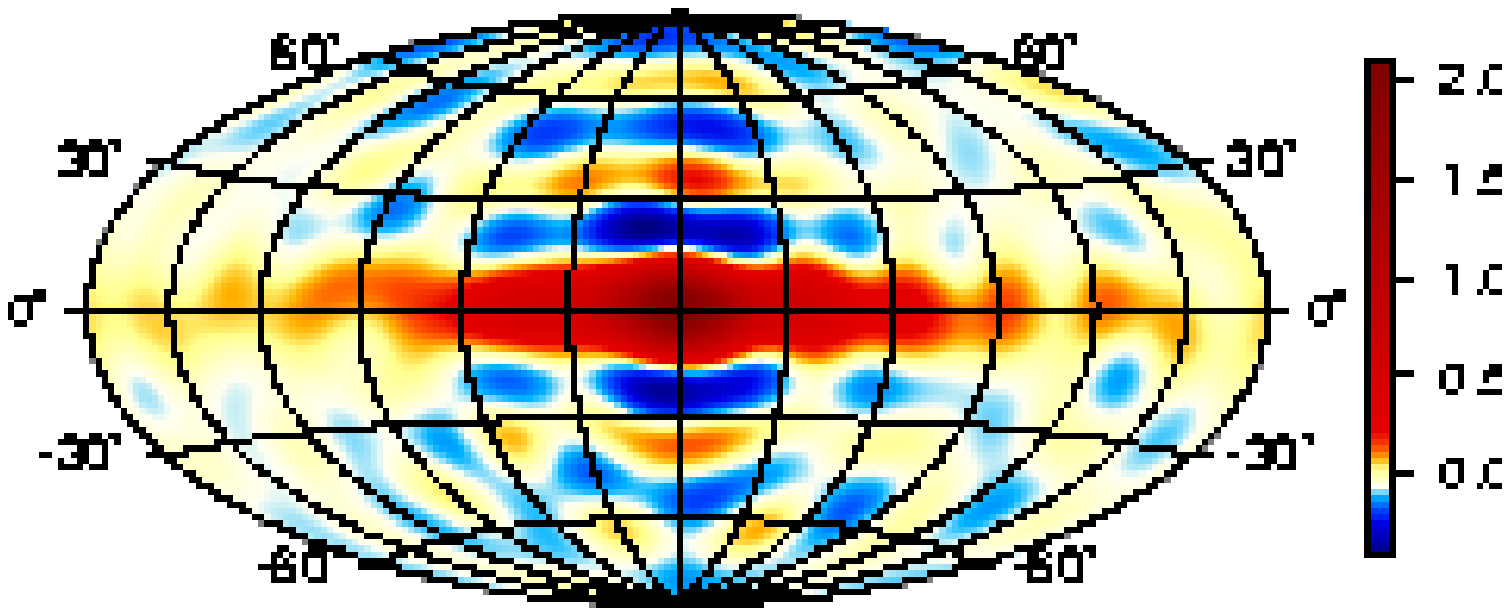}
\hspace*{0.0cm}
\includegraphics[width=5.8cm,clip,angle=0]{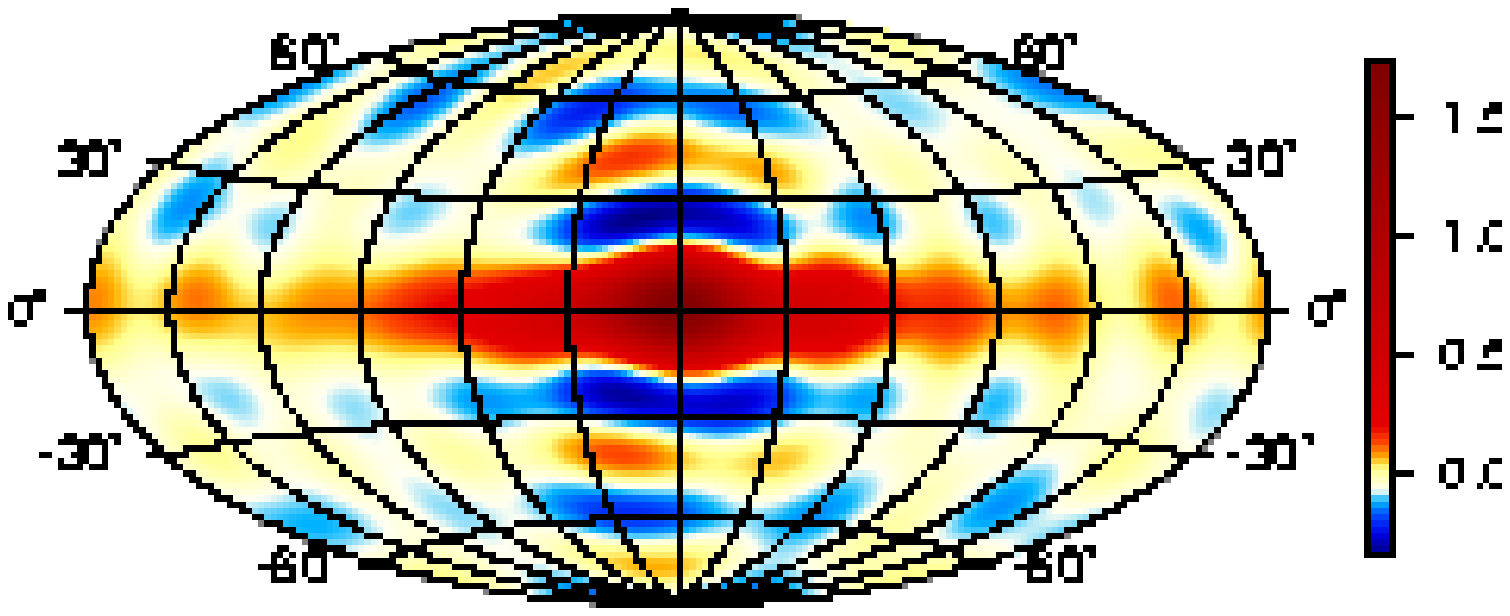}
\hspace*{0.0cm}
\includegraphics[width=5.8cm,clip,angle=0]{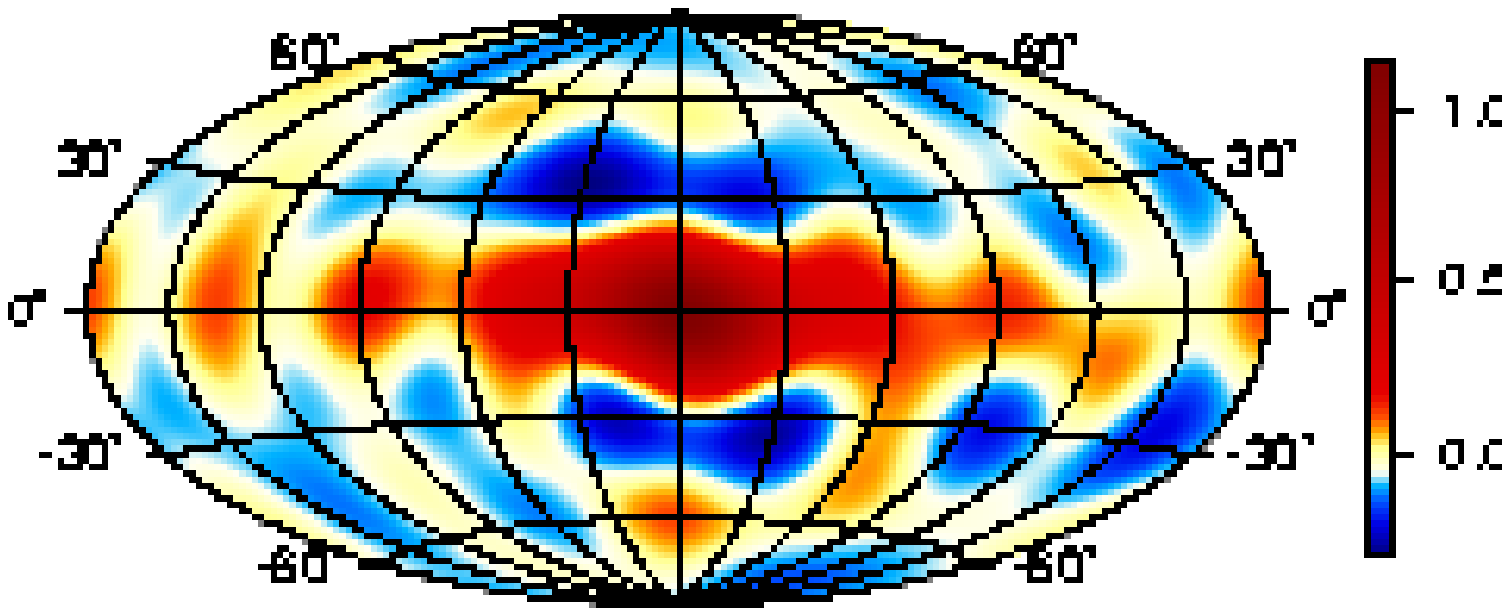}
\end{center}

\vspace*{-0.3cm}

    \caption{Dependence of the reconstructed results on the cutoff parameter 
      of singular values in absence of instrumental noises: 
      $w_{\rm cut}=10^{-6}$({\it left}), $10^{-4}$ ({\it middle}) and 
      $10^{-2}$({\it right}). }
    \label{fig:skymap_noisefree2}
\end{figure}

Fig.~\ref{fig:skymap_noisefree1} shows the intensity distribution of 
reconstructed skymap in the Galactic coordinate system. 
To obtain the skymap, we first create the annual modulation data 
of correlation signals $C_{AE}(f;t)$, $C_{AT}(f;t)$ and $C_{ET}(f;t)$ 
by convolving the original skymap 
$S_h(f,\mathbf{\Omega})=P(\mathbf{\Omega})$ with antenna pattern 
functions in the ecliptic coordinate system. Here the all-sky integral 
of $P(\mathbf{\Omega})$ is normalized 
to unity, i.e., $\int d\mathbf{\Omega} P(\mathbf{\Omega})=1$. 
Dividing the one-year data of correlation signals into 
the 32 sections ($N=32$), the vector $\mathbf{c}$ is then constructed.  
Combining it with the pseudo-inverse matrix of $\mathbf{A}$, 
we evaluate the expression (\ref{eq:approx}). The resultant least-squares 
solution of $p_{\ell m}$ is given in the ecliptic frame and we finally 
transform it into the skymap in the Galactic coordinate system. 
Note that in step calculating the matrix $\mathbf{A}$, the 
SPHEREPACK 3.1 package was used to compute the spherical harmonic 
coefficients of time-dependent 
antenna pattern functions and the multipoles higher than $\ell=17$ were 
discarded (i.e., $\ell_{\rm max}=16$).

The reconstruction result in Fig. \ref{fig:skymap_noisefree1} shows that 
due to the incomplete reconstruction,  the total intensity diminishes and 
the distribution is coarse-grained, together with fake pattern. 
Nevertheless, the disk-like structure of the intensity 
distribution is clearly seen. Compared to the result 
in the low-frequency band (paper II), 
the angular resolution is greatly improved.

It should be noticed that the present reconstruction result crucially 
depends on the additional parameter, i.e., the cutoff of singular 
values of the matrix $\mathbf{A}$, which we call $w_{\rm cut}$. 
For a given $w_{\rm cut}$, the pseudo-inverse matrix $A^{+}$ was constructed 
based on the expression (\ref{eq:pseudo-inverse}) using the 
singular-values larger than $w_{\rm cut}$.  
In Fig. \ref{fig:skymap_noisefree1}, we specifically 
set the cutoff parameter $w_{\rm cut}$ to $10^{-7}$. The cutoff value 
directly affects the resolution of reconstructed skymap. 
To show the significance of this, in Fig. \ref{fig:skymap_noisefree2}, 
reconstruction results of skymap with various cutoff are plotted. 
Clearly, larger cutoff value makes the angular resolution of skymap worse. 
It seems favorable to use the small value of $w_{\rm cut}$. However, 
as indicated in the expression (\ref{eq:pseudo-inverse}), 
the result using the small singular-value tends to be sensitively affected 
by the errors associated with numerical computation of pseudo-inverse matrix
and/or the instrumental noises. Hence, we need to seek an optimal value of 
$w_{\rm cut}$ to get a better reconstruction result. This will be 
discussed in next subsection. 

Now, we wish to quantify the quality of the reconstructed skymap and  
discuss the validity of present reconstruction scheme. To do this, 
we introduce the correlation parameter defined by 
\begin{eqnarray}
r_{\rm corr}(\ell_{\rm cut})\equiv 
\frac{\left\langle \left.S_h^{\rm(true)}(\mathbf{\Omega};\ell_{\rm cut})\right|S_h^{\rm(reconst)}(\mathbf{\Omega})\right\rangle}
{\left[ \left\langle \left.S_h^{\rm(true)}(\mathbf{\Omega};\ell_{\rm cut})\right|S_h^{\rm(true)}(\mathbf{\Omega};\ell_{\rm cut})\right\rangle \,\, 
      \left\langle \left.S_h^{\rm(reconst)}(\mathbf{\Omega})\right|S_h^{\rm(reconst)}(\mathbf{\Omega})\right\rangle\right]^{1/2}},
\label{eq:def_r_corr}
\end{eqnarray}
where the operation $\langle A|B \rangle$ for the real functions 
$A(\mathbf{\Omega})=\sum_{\ell m} a_{\ell m} Y_{\ell m}$ and 
$B(\mathbf{\Omega})=\sum_{\ell m} b_{\ell m} Y_{\ell m}$ means 
\begin{eqnarray}
\langle A | B \rangle &\equiv& 
\int d\mathbf{\Omega}\,\,
A(\mathbf{\Omega})B(\mathbf{\Omega})
= \frac{1}{2}\,\,\sum_{\ell,m}
\left(a_{\ell m}^*b_{\ell m}+ a_{\ell m}b_{\ell m}^*\right).
\end{eqnarray}
In the expression (\ref{eq:def_r_corr}), while the quantity
$S_h^{\rm (reconst)}$ represents the reconstructed skymap 
from the least-squares solution (\ref{eq:approx}), 
the function $S_h^{\rm (true)}$ is the skymap 
of the true intensity distribution dropping the higher multipole 
moments with $\ell>\ell_{\rm cut}$. To be precise, this is defined by 
\begin{equation}
S_{h}^{\rm(true)}(\mathbf{\Omega};\ell_{\rm cut})
=\sum_{\ell=0}^{\ell_{\rm cut}}
\sum_{m=-\ell}^{\ell}\,\,p_{\ell m}^{\rm(true)}\,\,
Y_{\ell m}(\mathbf{\Omega}). 
\end{equation}
The correlation parameter $r_{\rm corr}(\ell_{\rm cut})$ 
quantifies the degree of similarity between the  reconstructed skymap 
and the true skymap subtracting the higher multipoles. Hence, 
evaluating $r_{\rm corr}(\ell_{\rm cut})$, one can 
characterize the angular resolution of the GWB skymap. 
In addition to this, we also introduce the 
averaged fractional error for the multipole coefficients 
$p_{\ell m}$, $\mbox{Err}\,[p_{\ell m}]$, 
given by  
\begin{eqnarray}
\mbox{Err}\,[p_{\ell m}]&\equiv& 
\left\{
\frac{1}{2\ell+1} \,\sum_{m=-\ell}^{\ell}
\left|\frac{p_{\ell m}^{\rm (reconst)}-p_{\ell m}^{\rm (true)}}
{p_{\ell m}^{\rm (true)}}\right|^2 
\right\}^{1/2}, 
\label{eq:def_Err_plm}
\end{eqnarray}
which quantifies the accuracy of the determination of 
multipole coefficients. With the two quantities 
$r_{\rm corr}(\ell_{\rm cut})$ and $\mbox{Err}\,[p_{\ell m}]$, 
the quality and the accuracy of the reconstructed skymap can be 
quantified.

\begin{figure}
\begin{center}
\includegraphics[height=6.8cm,clip]{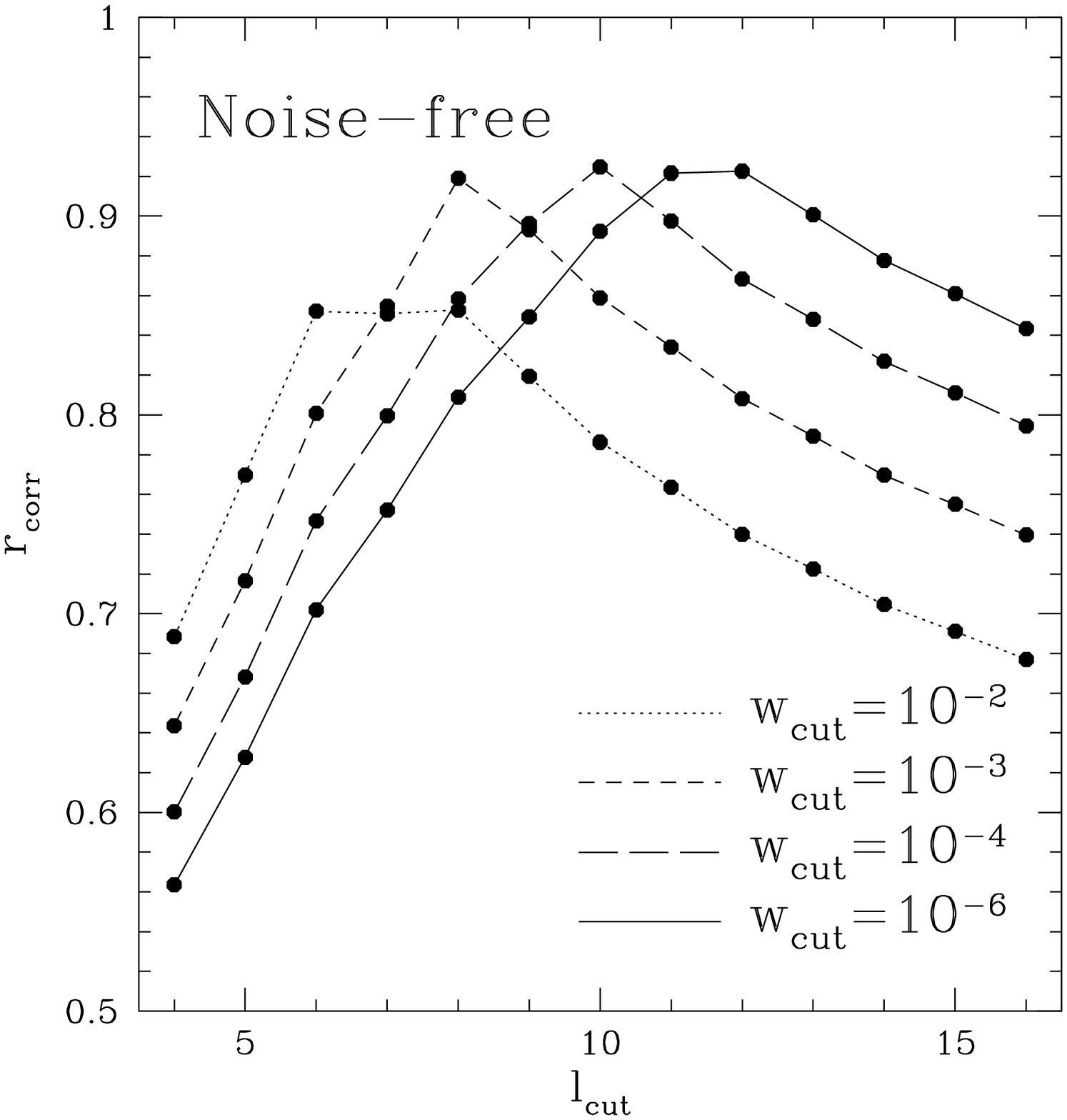}
\hspace*{1.0cm}
\includegraphics[height=6.8cm,clip]{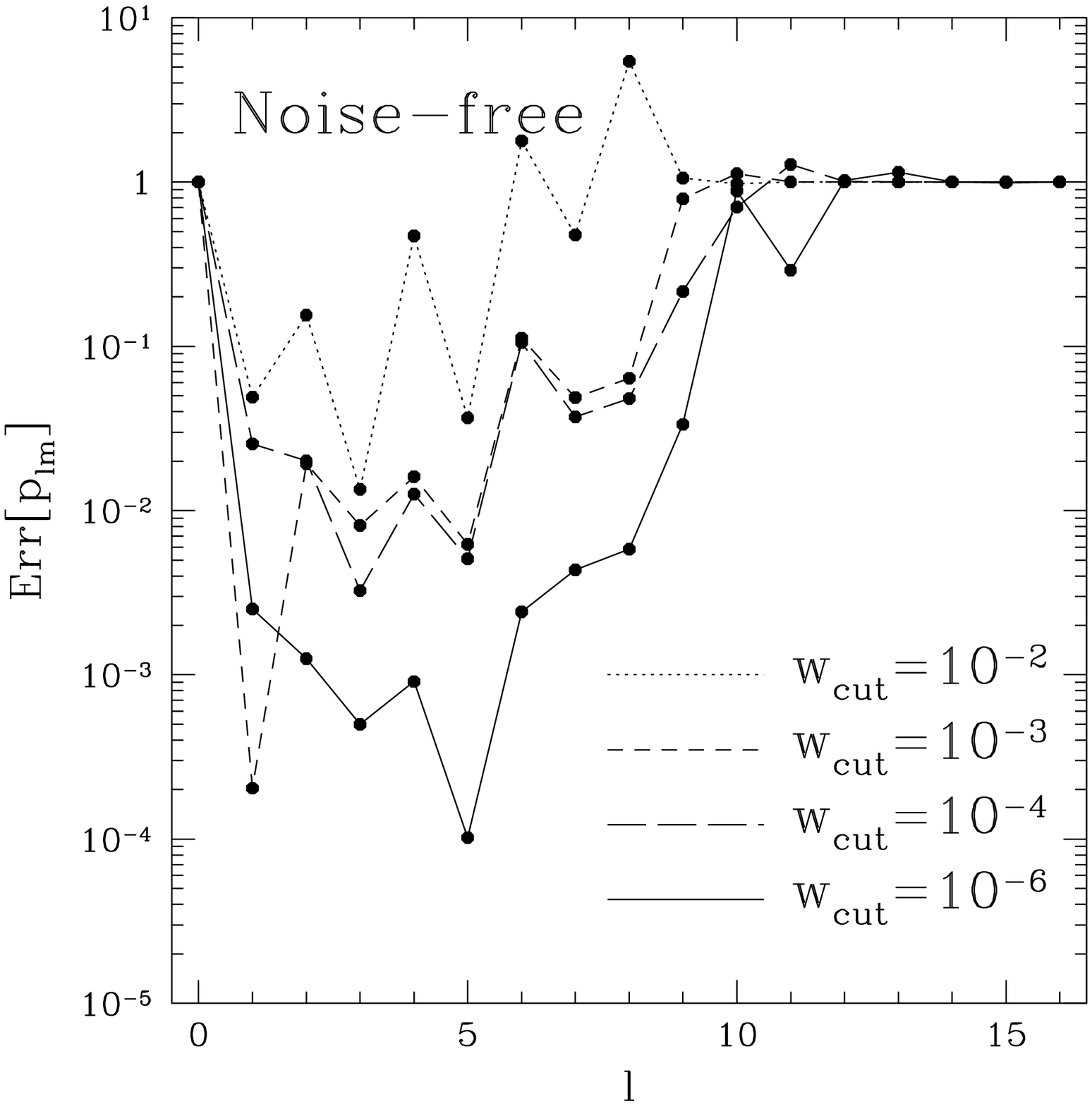}
\end{center}

\vspace*{-0.3cm}

    \caption{{\it Left}: Cross-correlation parameters $r_{\rm corr}$ 
      as function of 
      cutoff multipole $\ell_{\rm cut}$. {\it Right}:  
      averaged fractional error of $p_{\ell m}$, $\mbox{Err}\,[p_{\ell m}]$ 
      as function of multipole, $\ell$ (see Eq.(\ref{eq:def_Err_plm})).}
     \label{fig:rcorr_fracerror_noisefree}
\end{figure}

Left panel of Fig. \ref{fig:rcorr_fracerror_noisefree} shows  
the correlation parameters for the reconstructed skymap 
with various cutoff values $w_{\rm cut}$.  For a fixed value $w_{\rm cut}$, 
the correlation parameter as function of truncation multipole 
$\ell_{\rm cut}$ has a single peak and the peak 
value is typically $r_{\rm corr}\sim0.92$. As decreasing the cutoff 
value $w_{\rm cut}$, the location of peak tends to 
be shifted to a larger $\ell_{\rm cut}$, indicating that the  
angular resolution becomes improved. 
For small cutoff value $w_{\rm cut}=10^{-6}$, 
reconstructed skymap is qualitatively similar to the true skymap 
dropping the higher multipoles $\ell>12$.

Right panel of Fig. \ref{fig:rcorr_fracerror_noisefree} 
shows the root-mean-square value of the averaged fractional error of 
$p_{\ell m}$, i.e., $\mbox{Err}\,[p_{\ell m}]$. 
As anticipated from left panel, the smaller value of $w_{\rm cut}$ 
tends to decrease the errors in each multipole coefficient. However, 
due to the low-resolution of the antenna pattern function, 
the reconstructed results for higher multipoles $\ell>10$ 
become almost vanishing. As a result, the fractional error 
$\mbox{Err}\,[p_{\ell m}]$ of $\ell>10$ approaches unity.   
It should be noticed that the fractional error also becomes unity 
for the monopole moment of GWB signal. This reflects the important 
properties that the cross-correlation of optimal TDI variables are 
{\it generally blind} to the monopole intensity for the $AT$ and $ET$ signals
and also to the dipole anisotropy for the $AE$ signal. 
It has been shown in paper I that these features 
generally hold irrespective of the observed frequency band. 
In this respect, map-making issue with LISA would be problematic in 
determining of the monopole intensity. Apart from this, 
right panel of Fig. \ref{fig:rcorr_fracerror_noisefree} implies that 
the present reconstruction scheme has a potential to provide 
a precise determination of multipoles. With the cutoff value 
$w_{\rm cut}=10^{-6}$, the accuracy can reach 
less than a few percent level. Of course, these are the 
outcome based on the idealistic situation free from the instrumental 
noises. In next subsection, the significance of the 
instrumental noises on the map-making problem will be clarified. 

\subsection{Noisy case}
\label{subsec:noisy}

\begin{figure}[t]
\begin{center}
\includegraphics[width=16.5cm,clip]{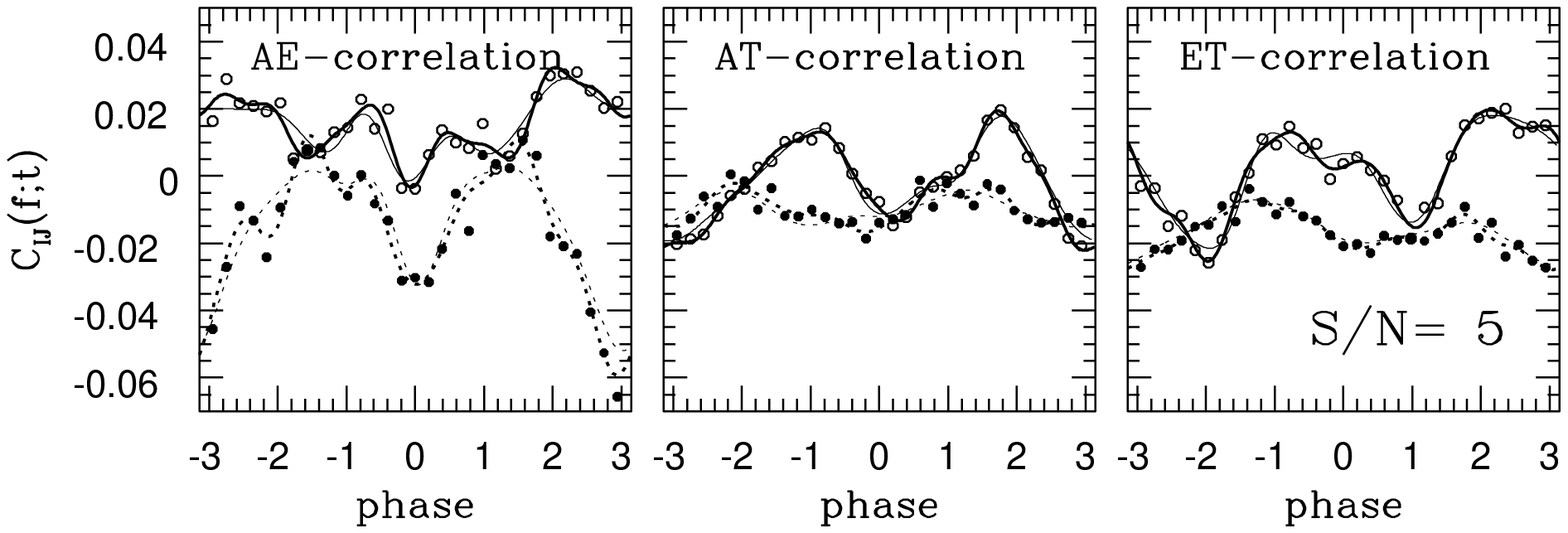}

\vspace*{0.5cm}

\includegraphics[width=16.5cm,clip]{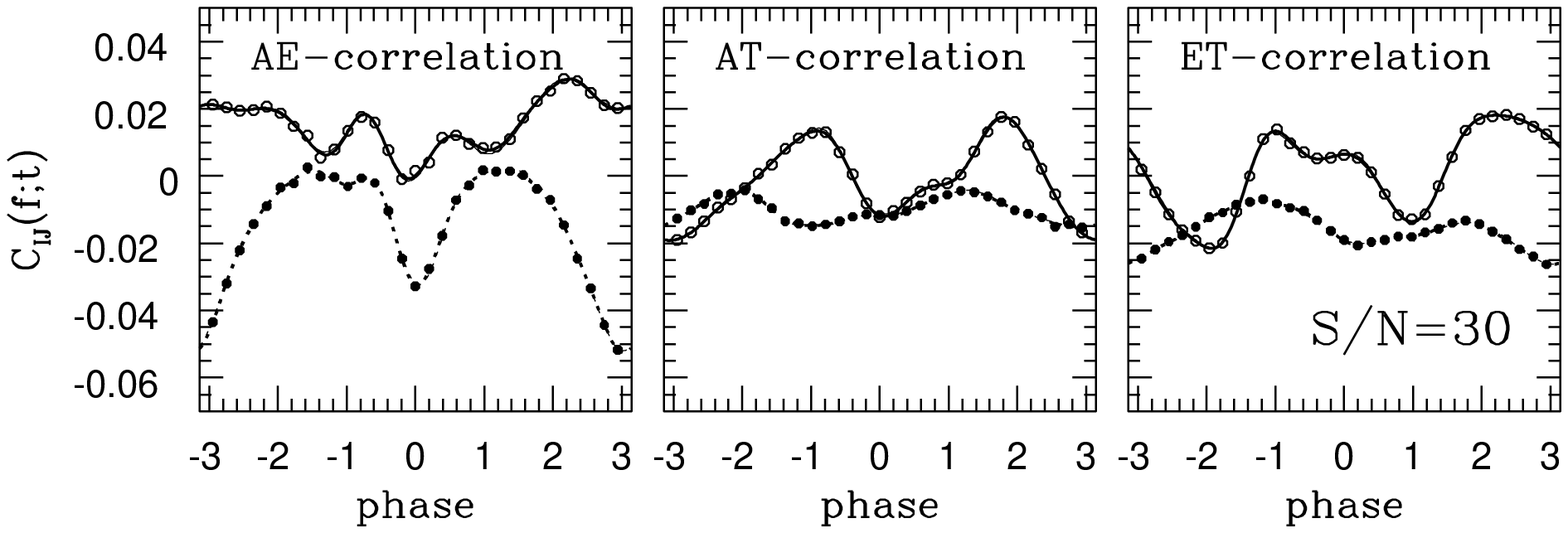}
\end{center}

\vspace*{-0.3cm}

    \caption{Annual modulation data of cross-correlation signals $C_{AE}$ 
      ({\it left}), $C_{AT}$ ({\it middle}) and $C_{ET}$ ({\it right}) 
      measured at the frequency $f=3f_*\simeq28.6$mHz. Top and bottom panels  
      show the results with signal-to-noise ratio    
      S/N$=5$ and $30$, respectively. In each panel, filled and open
      circles represent the real and the imaginary parts of the 
      noisy signals $\widehat{C}_{IJ}$ generated according to 
      Eq.~(\ref{eq:random_data}). While the continuous thin lines 
      indicates the correlation signals free from the detector noise, 
      thick lines indicate the fitted result of the noisy signals to 
      the harmonic functions (\ref{eq:harmonic_func}). 
       }
     \label{fig:annual_modulation}
\end{figure}

 \begin{figure}[t]
\begin{center}
\includegraphics[width=6.5cm,clip]{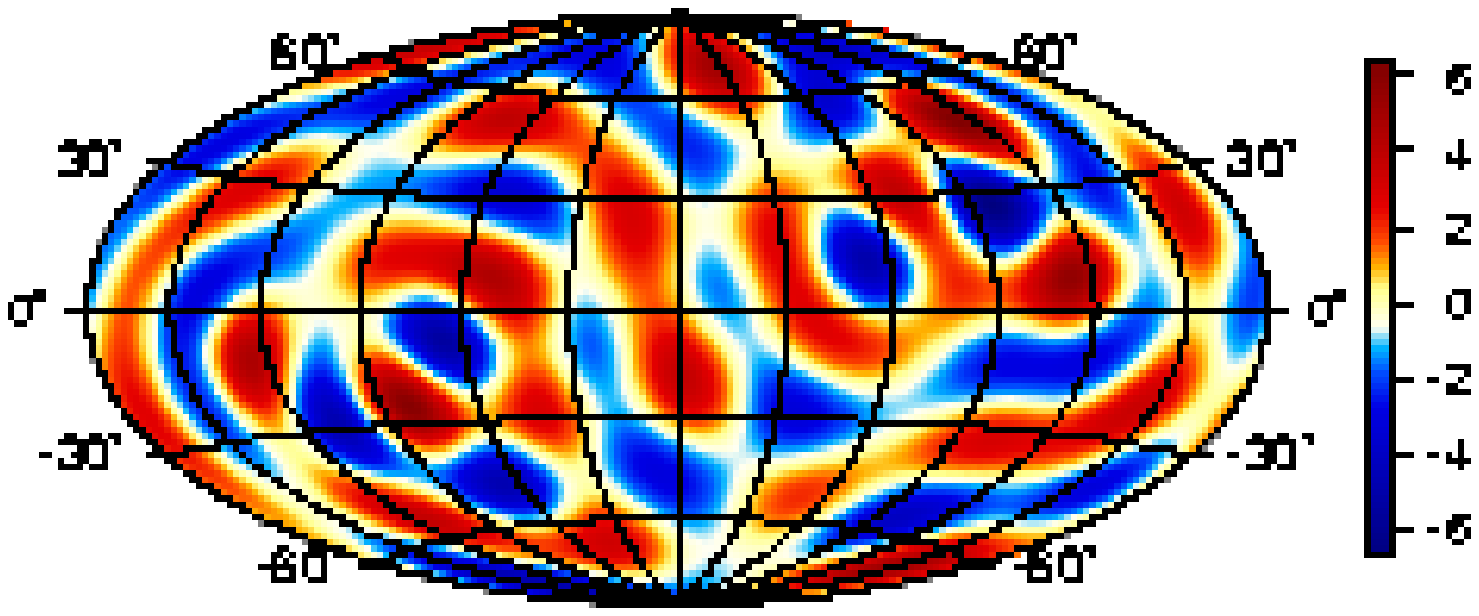}
\hspace*{0.5cm}
\includegraphics[width=6.5cm,clip]{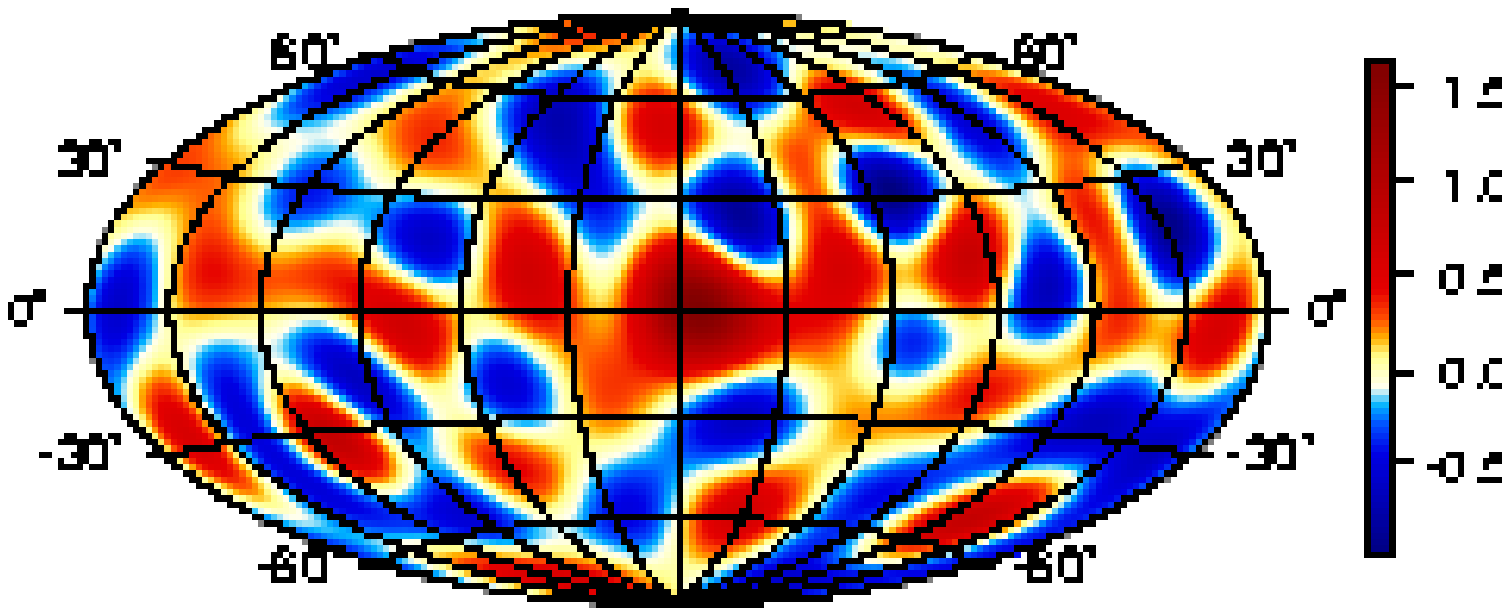}

\vspace*{0.4cm}

\includegraphics[width=6.5cm,clip]{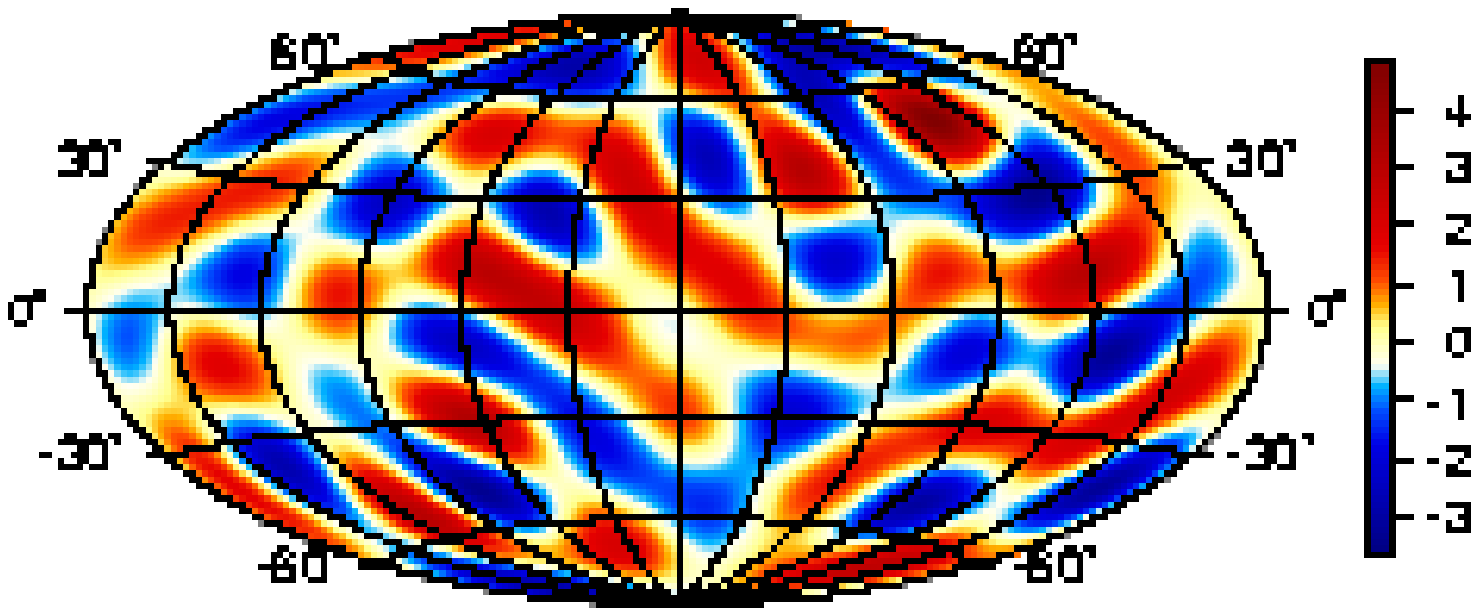}
\hspace*{0.5cm}
\includegraphics[width=6.5cm,clip]{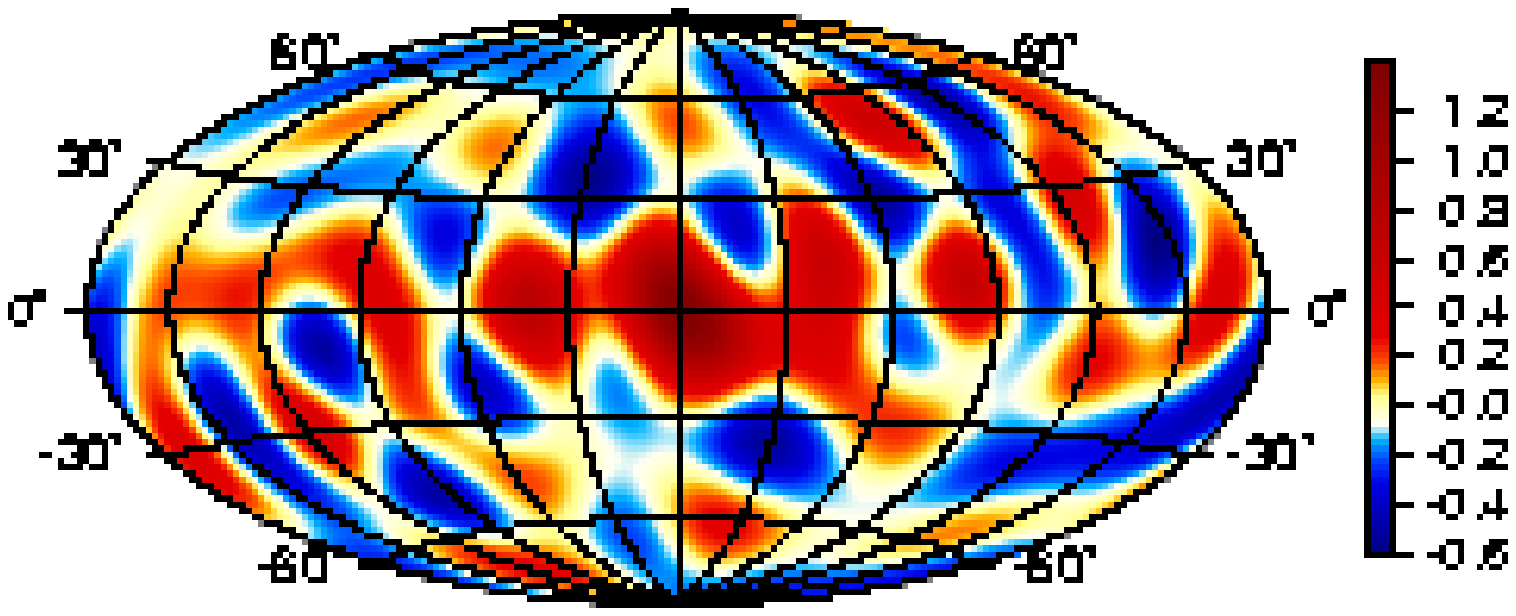}

\vspace*{0.4cm}

\includegraphics[width=6.5cm,clip]{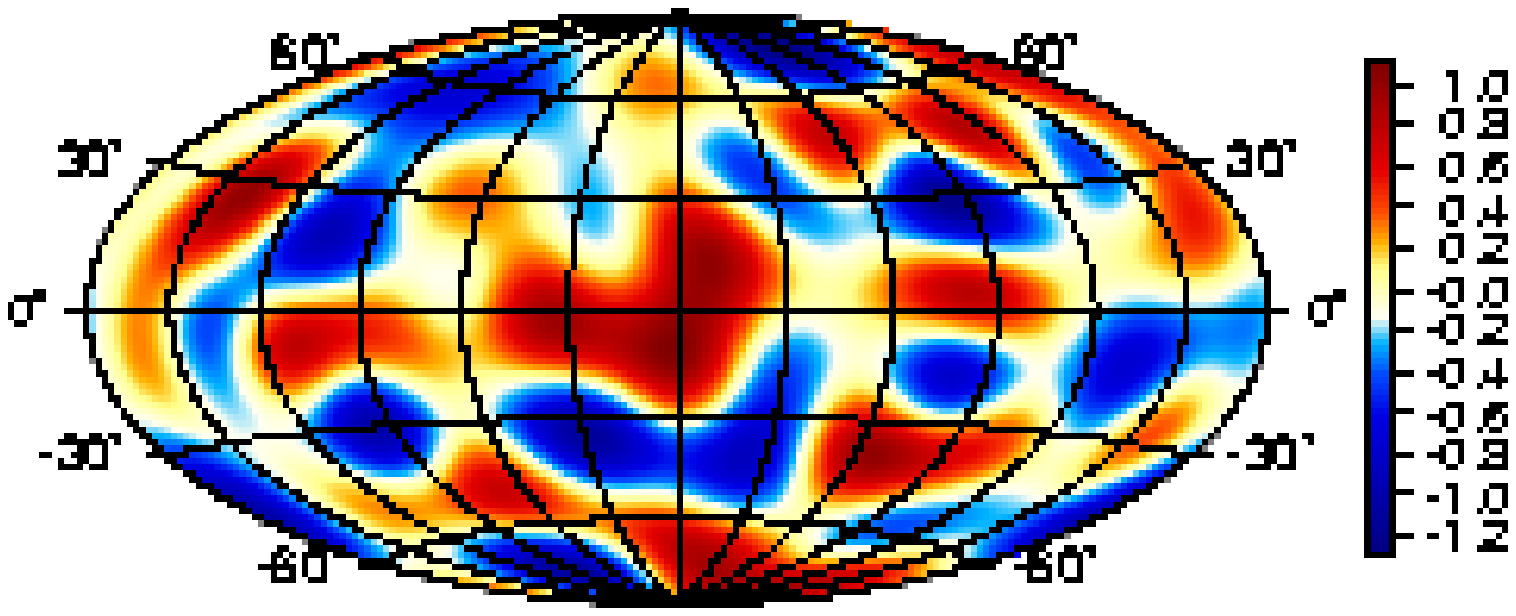}
\hspace*{0.5cm}
\includegraphics[width=6.5cm,clip]{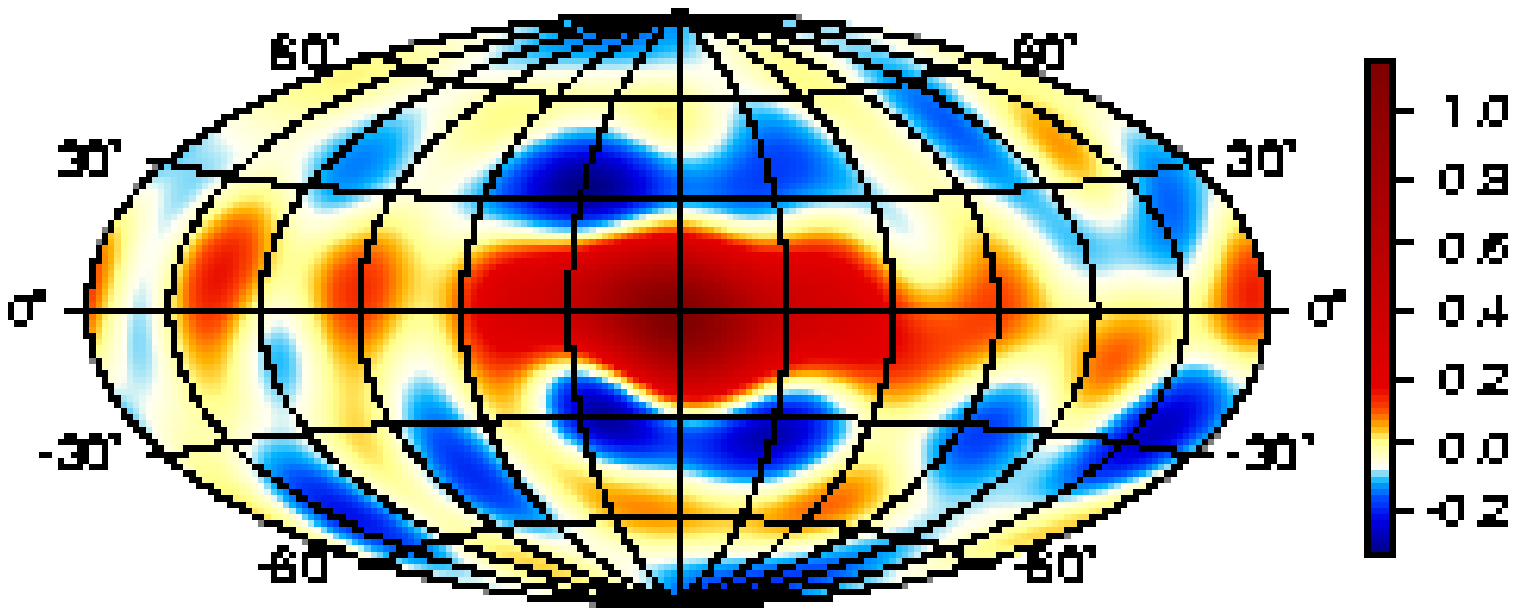}

\vspace*{0.4cm}

\includegraphics[width=6.5cm,clip]{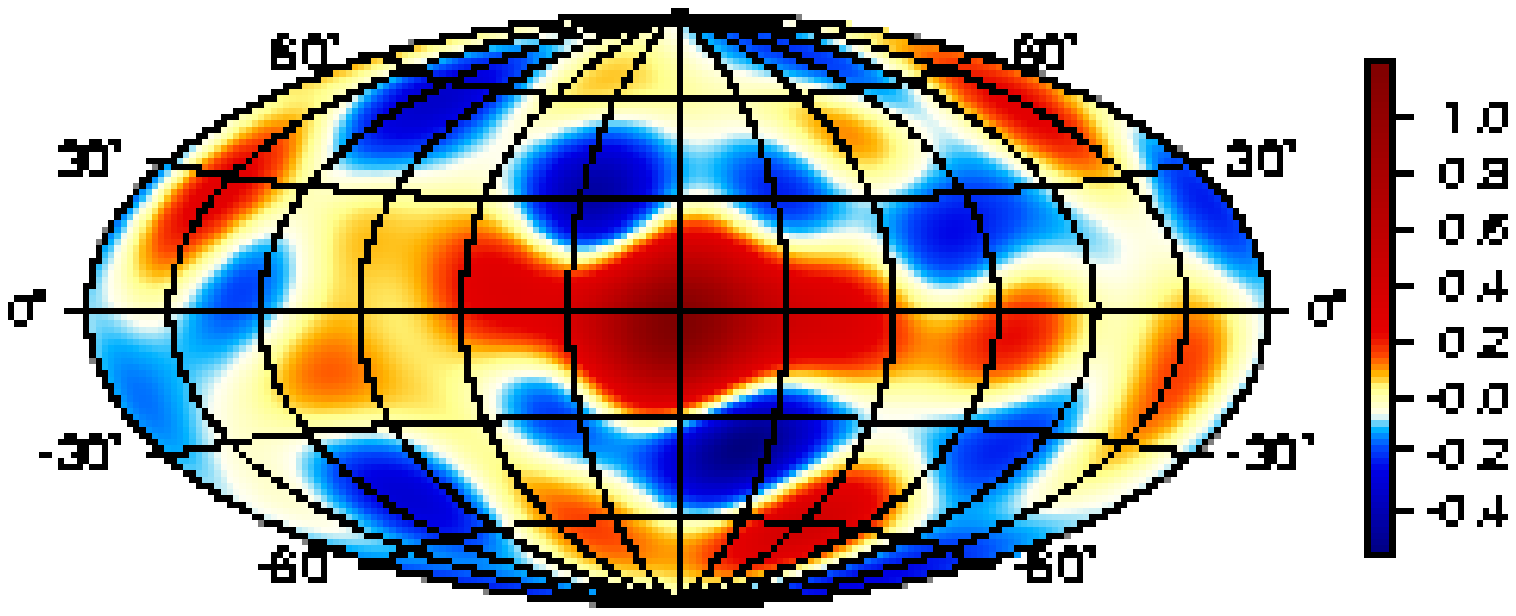}
\hspace*{0.5cm}
\includegraphics[width=6.5cm,clip]{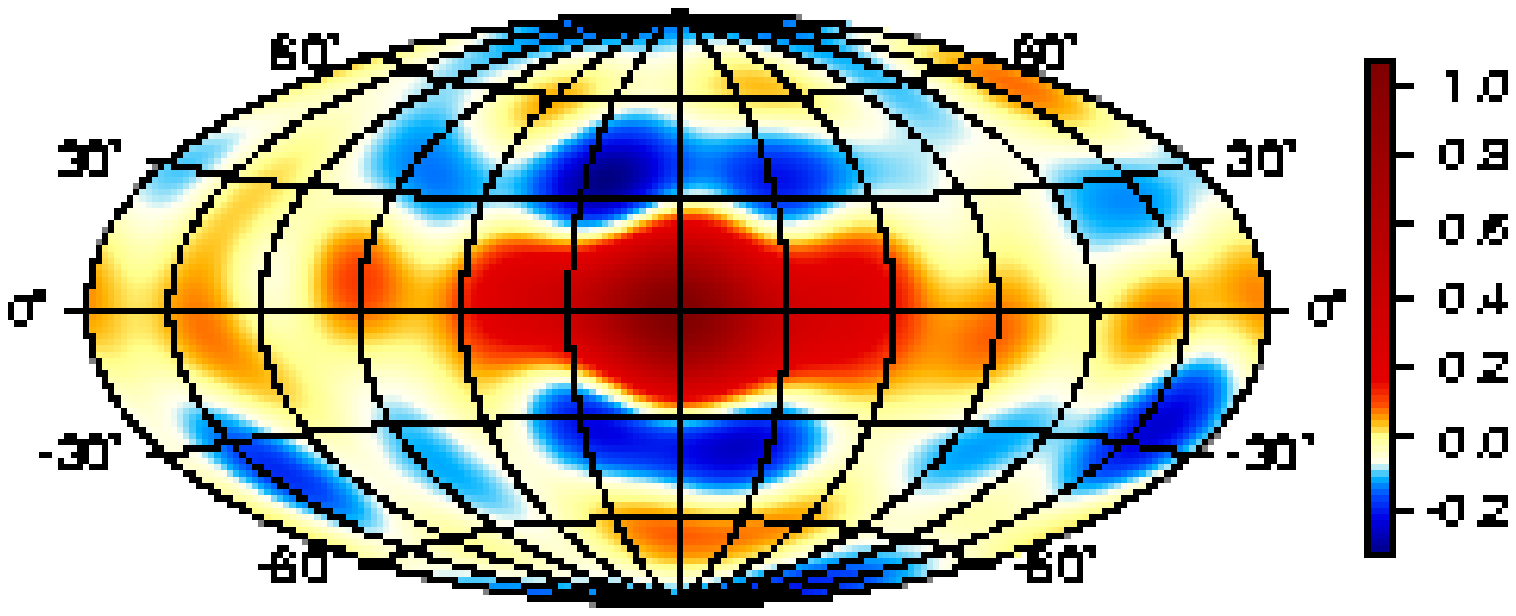}
\end{center}

\vspace*{-0.3cm}

\caption{Reconstructed skymap from the noisy data 
  (see Fig.~\ref{fig:annual_modulation}). 
  Left and right panels respectively show the results from the 
  S/N=5 and 30 noisy data. 
      From top to bottom, the cutoff parameter of singular 
      values were set to $w_{\rm cut}=10^{-3}$, $5\times10^{-3}$, 
      $10^{-2}$ and $5\times10^{-2}$.}
     \label{fig:noisy_skymap}
\end{figure}

In reality, 
the instrumental noises are additively mixed into the 
time-series data of gravitational-wave signals. 
While the cross-correlation variables  
$AE$, $AT$ and $ET$ themselves are free from the noise correlation, 
the correlation analysis with finite number of samples is inherently 
affected by the statistical fluctuations, among which the instrumental 
noises becomes the most dominant component in the weak-signal case. 
To discuss their  influences on the map-making problem, we first 
quantify the signal-to-noise 
ratio for the stochastic GWB. Based on the cross-correlation 
statistic with a suitable filter function, the signal-to-noise (S/N) 
ratio for the stochastic GWB in the specific frequency interval 
$f-\Delta f/2 \sim f+ \Delta f/2$ is defined by 
(e.g., \cite{Allen:1997ad,Kudoh:2005as}):
\begin{equation}
\left(\frac{S}{N}\right)\equiv (T_{\rm obs}\,\Delta f)^{1/2}\,
\left[ \sum_{(I,J)}\,\,
\frac{\overline{\,\,|C_{IJ}(f;t)|^2\,\,}}{N_I(f)N_J(f)}\right]^{1/2}
~; \quad
\overline{\,\,|C_{IJ}(f;t)|^2} \equiv \frac{1}{T_{\rm orbit}}
\int_0^{T_{\rm orbit}} dt\,\,|C_{IJ}(f;t)|^2,
\label{eq:def_SNR}
\end{equation}
where the summation $(I,J)$ represents the sum over the whole 
cross-correlation data (i.e., $AE$, $AT$ and $ET$). 
The quantities $T_{\rm orbit}$ and $T_{\rm obs}$ respectively denote 
the orbital period of LISA corresponding to the one year and 
the total observation time longer than the orbital period. 
Below, we specifically 
set the parameters to $T_{\rm obs}=10^8$sec and $\Delta f=10^{-3}$Hz. 
The function $N_{I}(f)$ represents the noise spectral density for the 
output data $I$. The explicit functional form of the noise spectrum 
is given in Appendix \ref{appendix:antenna_pattern},  
together with the specific parameters of the 
proof mass and the optical-path noises for LISA.

We are specifically concerned with the qualitative change of the 
reconstructed skymap in the presence of instrumental noises. 
For this purpose, rather than performing a large-scale extensive simulation 
that mimics a realistic signal processing 
\cite{Benacquista:2003th,Edlund:2005ye,Timpano:2005gm}, 
we here perform a very simple simulation 
in which the random noises are added by hand to the (noise-free) 
correlation signals computed in previous subsection. To be precise, 
for $i$-th sub-sectional data of the totally $N=32$ correlation signals 
$C_{IJ}$, we generate the time-series random data $\widehat{C}_{IJ}$ as 
\begin{eqnarray}
\widehat{C}_{IJ}(f;t_i) =C_{IJ}(f;t_i) + 
\left\{\frac{N_I(f)N_J(f)}{(T_{\rm obs}/N)\,
\Delta f}\right\}^{1/2}\,
\frac{\widehat{\xi}_1+i\,\widehat{\xi}_2}{\sqrt{2}}, \quad 
(i=1,\cdots,N)
\label{eq:random_data}
\end{eqnarray}
where the variables $\hat{\xi}_i$ are the Gaussian random variables 
with zero mean and unit variance, i.e., 
$\langle\hat{\xi}_i\rangle=0$ and $\langle\hat{\xi}_i^2\rangle=1$.  
In the above expression, the first term in right-hand-side is the 
theoretical time-modulation signals used in previous subsection, 
where the amplitude of GWB signal,  
$S_h(f,\mathbf{\Omega})=A\,\cdot\,P(\mathbf{\Omega})$. On the other hand,  
the second term represents the randomness arising from 
the instrumental noises, whose amplitude is estimated based on 
the S/N ratio in weak-signal limit.
%

Fig. \ref{fig:annual_modulation} shows the time-series data of 
cross-correlation signals for Galactic GWB in the case of S/N$=5$ 
({\it top}) and S/N$=30$ ({\it bottom}). The open and the filled circles 
represent the real and the imaginary parts of the noisy data 
$\widehat{C}_{IJ}$, respectively. Note that in plotting the data, 
we set $A=1$ so that the all-sky integral of GWB spectrum is normalized to 
unity and the amplitude of the random noises is appropriately 
rescaled according to the S/N values. As it is clear, direct use of 
the raw noisy data makes the quality of the reconstructed image worse. 
Hence, we tried to fit the noisy data $\widehat{C}_{IJ}$ 
to the harmonic functions $f_{IJ}(t)$: 
\begin{equation}
f_{IJ}(t)=\sum_{k=-k_{\rm max}}^{k_{\rm max}}\,\,c_k\,\,
e^{ik\,\omega_{\rm orbit}\,t}\,\,;\quad \omega_{\rm orbit}=2\pi/T_{\rm orbit}.
\label{eq:harmonic_func}
\end{equation}
The resultant fitting functions were then used to perform the 
reconstruction scheme. 
In Fig. \ref{fig:annual_modulation}, the thick-solid and the thick-dashed 
lines are the results fitted to the harmonic function 
(\ref{eq:harmonic_func}) with $k_{\rm max}=8$. For comparison, 
we also plot the continuous thin lines as the noise-free correlation signals.

Fig. \ref{fig:noisy_skymap} shows the reconstructed results from the 
noisy data for the various cutoff parameters of the singular-values,  
$w_{\rm cut}$: $w_{\rm cut}=10^{-3}$, $5\times10^{-3}$, $10^{-2}$ and 
$5\times10^{-2}$ from top to bottom. 
The left panel plots the results in the S/N$=5$ cases, 
while the right panel shows the skymap in the S/N$=30$ cases. 
The skymap with small cutoff value $w_{\rm cut}=10^{-3}$ is affected by the 
instrumental noises and the resultant intensity map show featureless fake 
patterns. As increasing the cutoff parameter, 
fake intensity pattern gradually disappears and the strong intensity peak 
seen in the original GWB map becomes prominent. The quality of the 
resultant skymap with large $w_{\rm cut}$ depends on the signal-to-noise 
ratio, S/N. For the cutoff parameter $w_{\rm cut}=10^{-2}$, 
the resultant skymap in the S/N=30 case is very similar to the one in 
the noise-free result with 
$w_{\rm cut}=10^{-2}$ (right panel of Fig. \ref{fig:skymap_noisefree2}), 
while the fake intensity image is still dominant in the S/N=5 case.

The influence of instrumental noises shown in Fig. \ref{fig:noisy_skymap} 
may be easily deduced 
from the expression  (\ref{eq:approx}) with (\ref{eq:pseudo-inverse}). 
In the presence of the noise, the vector $\mathbf{c}$ which represents 
the correlation signals additionally contains the noise term $\mathbf{n}$ 
and one can write it as $\mathbf{c}=\mathbf{A}\cdot\mathbf{p}+\mathbf{n}$.  
Then, the least-squares approximation (\ref{eq:approx}) leads to the unwanted 
term $\mathbf{A}^+\cdot\mathbf{n}$. Since the pseudo-inverse matrix 
$\mathbf{A}^+$ contains the reciprocal of the singular values, 
$\{w_i^{-1}\}$,  this additional contribution can become dominant 
and affect the final reconstructed 
result unless introducing a larger cutoff value $w_{\rm cut}$. 
Therefore, in order to reduce the influence of the noise term, the cutoff 
parameter should be, at least, set to $\sim$(S/N)$^{-1}$.

\begin{figure}[t]
\begin{center}
\includegraphics[height=6.8cm,clip]{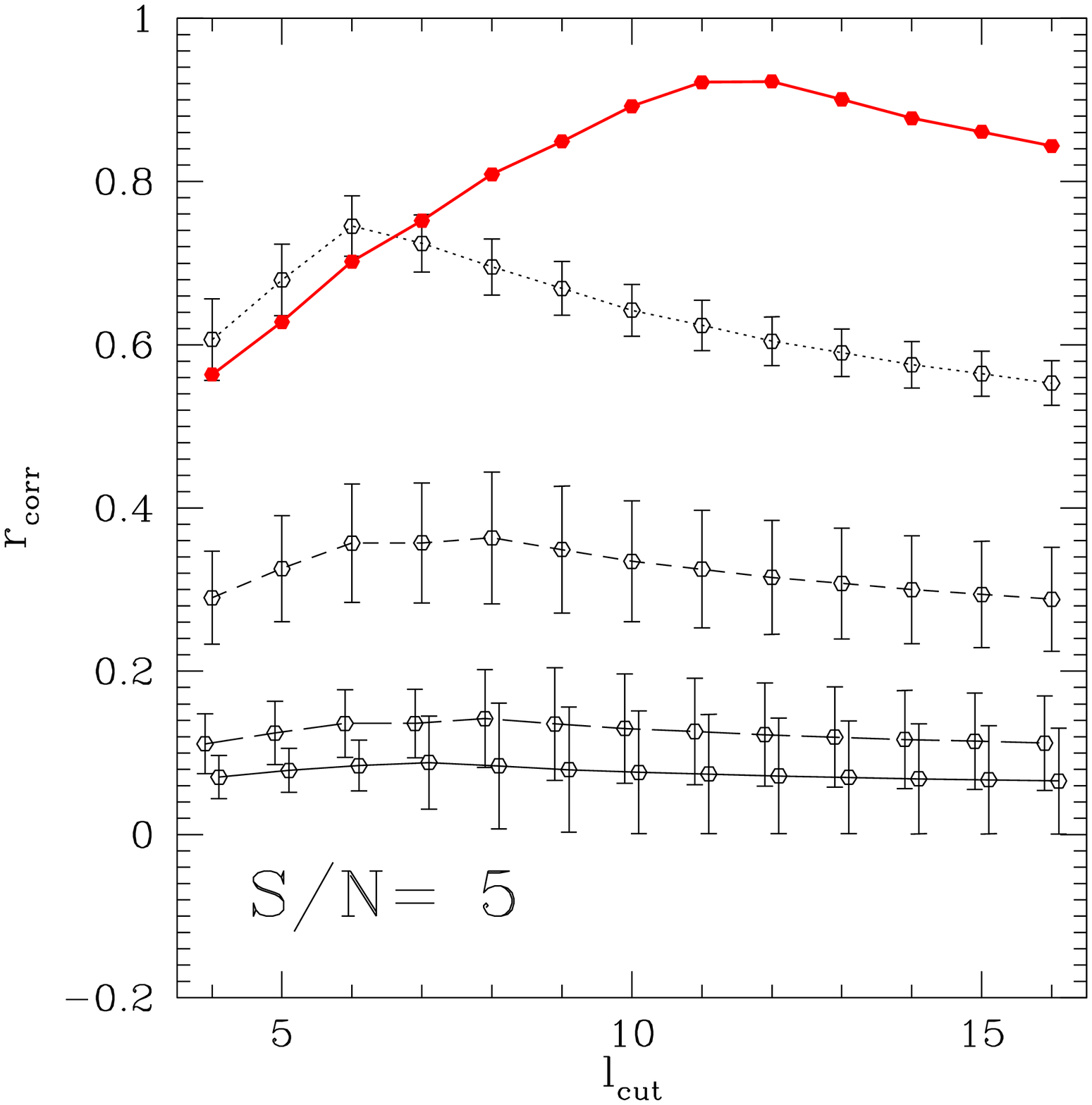}
\hspace*{0.8cm}
\includegraphics[height=6.8cm,clip]{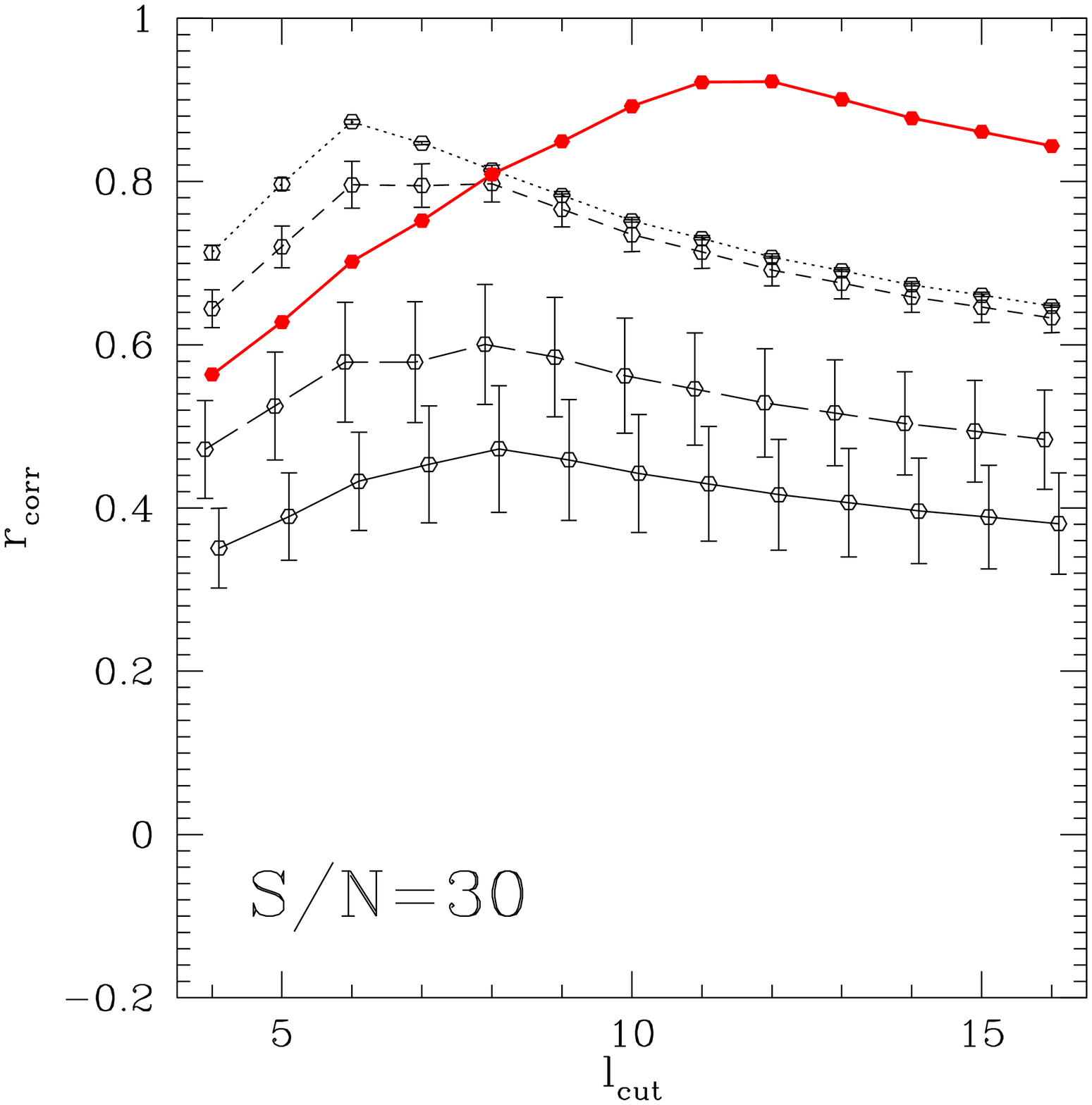}
\end{center}

\vspace*{-0.3cm}

\caption{Cross-correlation parameters of reconstructed skymap in presence of 
  the instrumental noises with S/N=5 ({\it left}) and S/N=30 ({\it right}). 
  In both panels, the cutoff parameters of the singular-values of the matrix 
  $\mathbf{A}$ were set to $w_{\rm cut}=10^{-3}$ ({\it solid}), 
  $5\times10^{-3}$ ({\it long-dashed}), $10^{-2}$ ({\it short-dashed}) 
  and $5\times10^{-2}$ ({\it dotted}), from bottom to top.  
  The error bars are estimated from the sample variation among $100$ 
  realizations.  For 
  comparison, the cross-correlation parameter in the noise-free case 
  ($w_{\rm cut}=10^{-6}$) is also plotted in thick solid line.   
}
     \label{fig:rcorr_noisy}
\end{figure}

\begin{figure}[t]
\begin{center}
\includegraphics[width=6.8cm,clip]{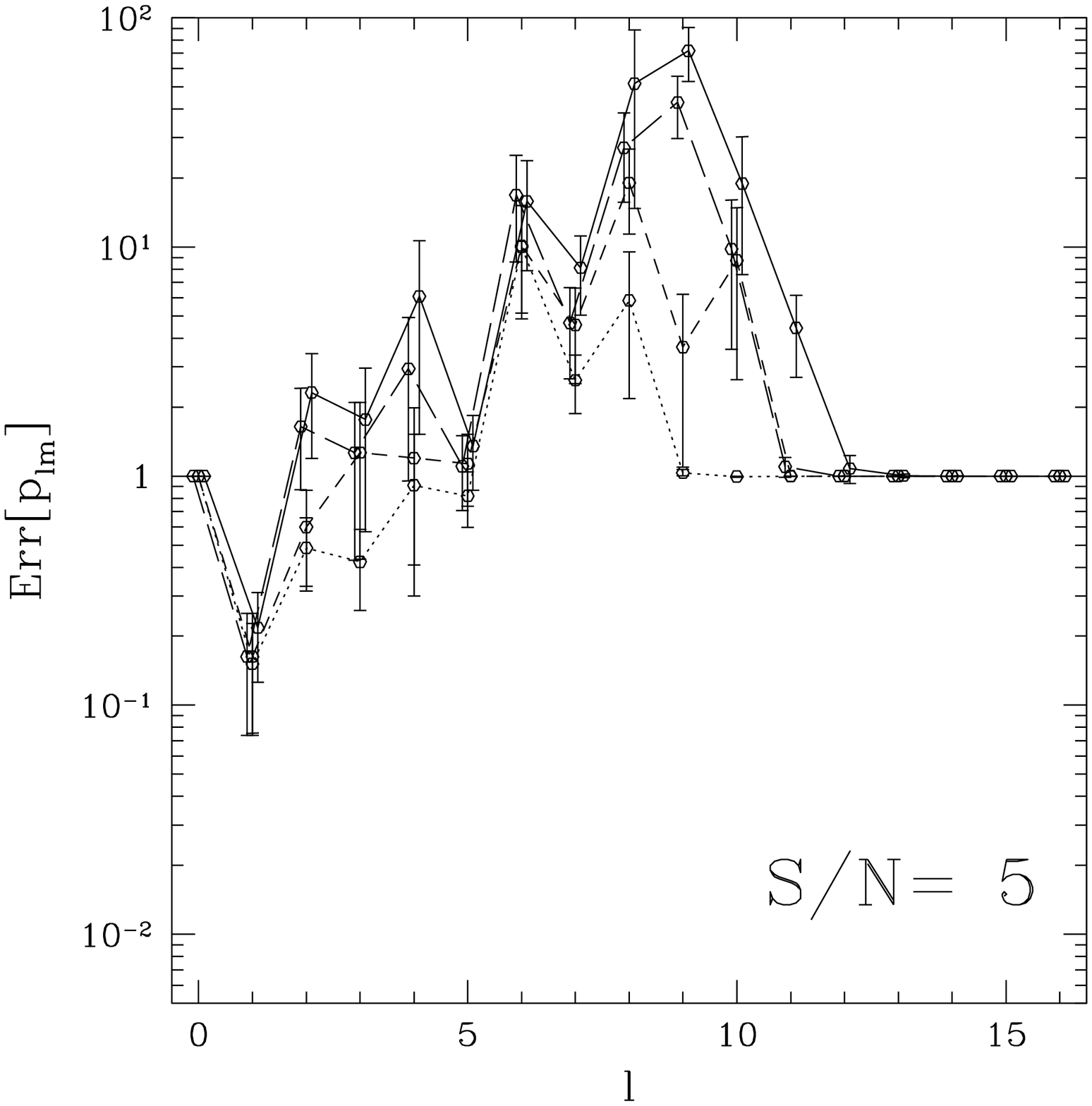}
\hspace*{0.8cm}
\includegraphics[width=6.8cm,clip]{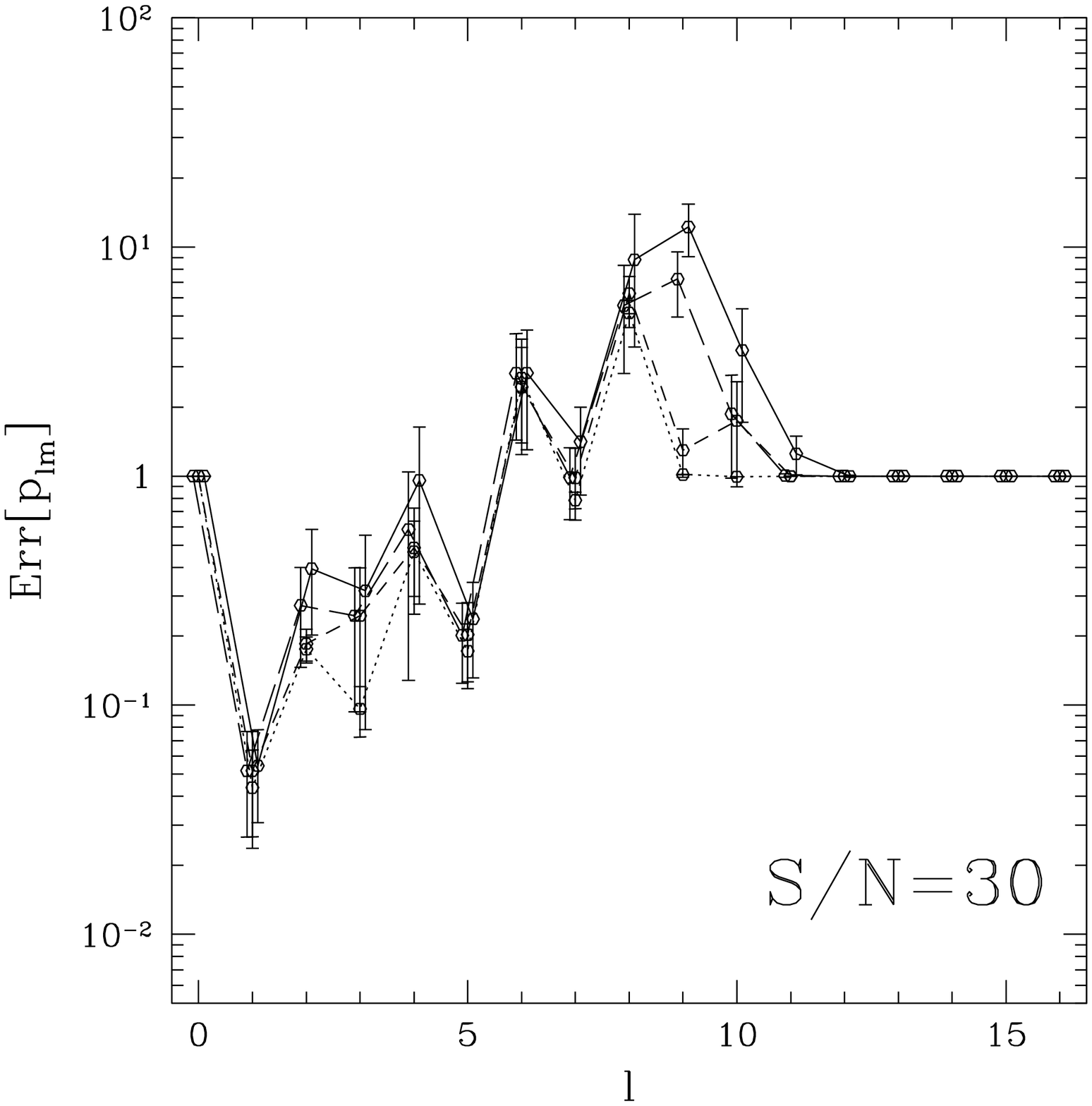}

\vspace*{-0.3cm}

\end{center}
    \caption{Averaged fractional error of $p_{\ell m}$, 
      $\mbox{Err}\,[p_{\ell m}]$ 
      as function of $\ell$ in the presence of noises. The left and right 
      panels represent the results with S/N=5 and 30, respectively. In 
      both panels, the cutoff parameters $w_{\rm cut}$
      were chosen as $10^{-3}$ ({\it solid}), 
      $5\times10^{-3}$ ({\it long-dashed}), $10^{-2}$ ({\it short-dashed}) 
      and $5\times10^{-2}$ ({\it dotted}).  The error bars indicate 
      the $1\sigma$ variation among $100$ realizations.   
}
     \label{fig:fracerror_noisy}
\end{figure}

In Fig. \ref{fig:rcorr_noisy}, quality of the reconstructed skymap 
is quantified by evaluating the correlation parameters $r_{\rm corr}$. Also, 
the averaged fractional errors $\mbox{Err}\,[p_{\ell m}]$ are computed 
and are presented in Fig. \ref{fig:fracerror_noisy}  
for various cutoff values. 
In both figures, the error bars indicate the $1\sigma$ 
variation among $100$ realizations. 
As anticipated from Fig. \ref{fig:noisy_skymap}, 
the quality of final skymap is significantly degraded in 
the case of the low S/N data.  
The result with  a smaller cutoff value $w_{\rm cut}$ has a little 
correlation with the true skymap due to many fake patterns, 
which is mainly attributed to the errors in the higher multipoles $\ell>8$. 
For the high-signal-to-noise ratio S/N=30, the situation is somehow 
improved. With a cutoff value around 
$w_{\rm cut}\sim(\mbox{S/N})^{-1}$, 
the quality of reconstructed image becomes similar to the true map 
with $\ell\lesssim6-7$. The fractional error in each multipole 
coefficient reaches $\sim10\%$, although it seems still miserable 
compared with the noise-free result.

\begin{figure}[ht]
\begin{center}
\includegraphics[width=8cm,clip]{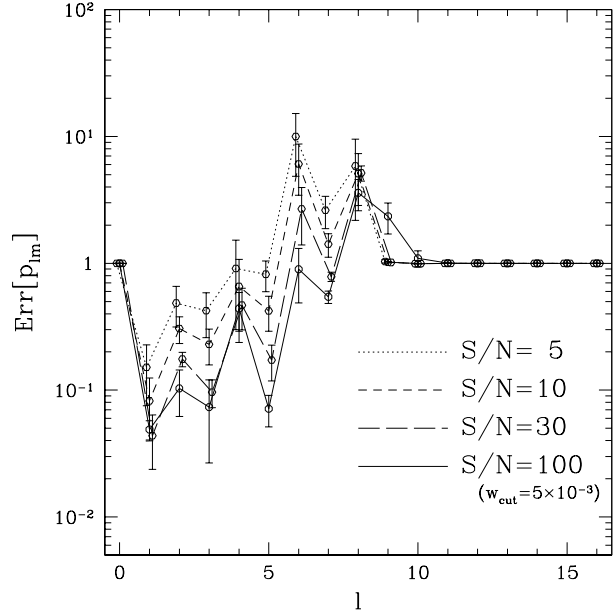}
\end{center}

\vspace*{-0.3cm}

    \caption{Same as in Fig.~\ref{fig:fracerror_noisy}, but 
      the dependence of signal-to-noise ratio is shown. 
      The solid, long-dashed, short-dashed and dotted lines 
      respectively show the results with S/N=100, 30, 10 and 5. 
      In each case, except for S/N=100, the cutoff parameter 
      $w_{\rm cut}$ was set to $5\times10^{-2}$. In the case of the 
      high signal-to-noise ratio S/N=100, 
      fractional error can be further reduced if one adopt  
      a smaller cutoff value. In this plot, the result 
      with $w_{\rm cut}=5\times10^{-3}$ is shown.  
}
     \label{fig:fracerror_noisy2}
\end{figure}

Finally, accuracy of the reconstructed 
multipoles achieved with a given signal-to-noise ratio S/N is 
shown in Fig.~\ref{fig:fracerror_noisy2}. 
Except for S/N=100, the cutoff value was all fixed to 
$w_{\rm cut}=5\times10^{-2}$. 
For high signal-to-noise ratio S/N=100, 
the accuracy is further improved adopting the small cutoff parameter, 
$w_{\rm cut}=5\times10^{-3}$. 
In Fig.~\ref{fig:fracerror_noisy2}, 
apart from the monopole moment and the higher multipoles of $\ell>8$, 
there exist some multipoles that are still difficult to determine.  
Recalling from the similar trend 
found in the noise-free case 
(right panel of Fig.~\ref{fig:rcorr_fracerror_noisefree}), 
the non-uniformity of the accuracies might be attributed to the 
angular response of the antenna pattern functions at the 
frequency $f=3f_*$, not the character of the specific model of 
anisotropic GWB. This implies that 
the multi-frequency data analysis is important for the accurate 
determination of the multipole moments $p_{\ell m}$. 
Anyway, the quantitative estimation of $p_{\ell m}$ 
at few percent level seems very difficult even when S/N=100. 
This means that in contrast to the low-frequency case, 
the present reconstruction method might not be so 
powerful to determine each multipole moment of high-frequency skymap, 
but should be rather helpful to study the all-sky distribution of 
anisotropic GWBs in bird-eye view.

\section{Summary \& Discussion}
\label{sec:conclusion}

In this paper, we have discussed the reconstruction of 
a high-frequency skymap of gravitational-wave 
backgrounds with a space-based interferometer. Owing to the 
cross-correlation technique, 
we present a simple numerical scheme to reconstruct a skymap of 
anisotropic stochastic signals.  While the methodology presented here 
is basically the same one as described in paper II, i.e., least-squares 
solution of linear algebraic system, there are several distinctions  
in the reconstruction of the high-frequency skymap. 
First of all, the antenna pattern function becomes 
complicated and the analytic treatment based on the perturbative 
expansion of antenna pattern function is intractable. 
Hence, a full numerical treatment to reconstruct the higher multipole 
moments is necessary. On the other hand, 
the sensitivity to the high-frequency signals is improved 
and no multi-frequency data is needed for the reconstruction of 
GWB skymap. As a result, the skymap with LISA demonstrated in the 
Galactic GWB case yields a better angular resolution up to  
the multipoles $\ell\sim10$ in the absence of the instrumental noises. 
The presence of noises degrades the angular resolution 
and the multipole coefficients for GWB skymap includes a larger error.  
However, with the signal-to-noise  ratio S/N$>5$, the resultant 
skymap has angular resolution with multipoles $\ell\leq 6\sim7$, 
which gives a better quality compared to the one achieved by 
the low-frequency skymap.

Since the methodology presented here is still very primitive, there would 
be several extensions to improve the angular resolution of reconstructed 
skymap. Perhaps, one naive approach 
is the combination of the multi-frequency data set. With the multiple data, 
one effectively increases the signal-to-noise ratio. Also, 
the directional information for anisotropic signals would be 
obtained additionally, 
through the frequency-dependent angular response of antenna 
pattern functions. Another possible approach is to combine the parametric 
models of GWB distribution characterized by the finite number of 
parameters and to determine these parameters. With fewer parameter set, 
the present methodology tightly constrains the model parameters.

The reconstructed skymap obtained from the present scheme 
may be regarded as the first step of the synthesis imaging processing, 
like a {\it dirty map} in radio astronomy \cite{Thompson:1986}. 
Under a priori information about the source distribution and the 
reasonable assumptions, 
it would be CLEANed by using the iterative deconvolution algorithm 
(see Ref.\cite{Cornish:2003vj} for the application of CLEAN algorithm 
to the data analysis of gravitational waves).  
That is, with the compact GWB distribution coming from the 
nearby sources, one can create the synthesized GWB map convolving with the 
antenna pattern functions. We then subtract each compact component 
from the dirty map and repeat the procedure iteratively until all 
significant source structure has been removed. A CLEAN map 
is finally obtained from the residual intensity distribution by adding 
the removed GWB components with suitably smoothed sampling function. 
This CLEAN map would be useful and helpful to discriminate 
the GWBs of nearby origin and cosmological origin and even to  
identify the specific GWB signals.

Finally, the map-making problem considered in the paper is 
related to the tomographic reconstruction technique to resolve the 
distribution of the binaries \cite{Mohanty:2005ca}. 
While the present technique only 
relies on the amplitude of the signals, the tomographic approach 
fully takes account of the phase information. As a result, the 
angular resolution is greatly improved and the identification of 
signals becomes efficient even in the crowded samples. This may be 
a hint to improve the angular resolution of GWB skymap.

\begin{acknowledgments}
We thank Hideaki Kudoh, Yoshiaki Himemoto, Takashi Hiramatsu and 
Shun Saito for discussions and comments. We also thank Naoki Seto 
for careful reading of the manuscript and discussions. 
This work was supported by a Grant-in-Aid for Scientific 
Research from the Japan Society for the Promotion of Science 
(No.~18740132). 
\end{acknowledgments} 

\appendix

\section{Antenna pattern functions and 
  instrumental noises}
\label{appendix:antenna_pattern}

Here, we give an explicit expressions for antenna pattern functions for LISA
used in the main text. First recall the definition of the 
antenna pattern function (paper I, II, \cite{Cornish:2001hg}):
\begin{eqnarray}
    && \mathcal{F}_{IJ}(f, \mathbf{\Omega};\,t)=  
    e^{ i \, 2\pi f \,{\bf \Omega \cdot}(\textbf{x}_I -\textbf{x}_J) }
    \sum_{A=+,\times} 
    F_I^{A*}(  \mathbf{\Omega},f;\,t) F_J^A( \mathbf{\Omega},f;\,t)
    \cr
    && F_I^A(  \mathbf{\Omega},f;\,t)  = 
    {\bf D}_I( \mathbf{\Omega},f;\,t)\, \textbf{:} \,
    \textbf{e} ^A (\mathbf{\Omega}), 
\label{eq:def_antenna}
\end{eqnarray}
where $\textbf{e} ^A$ is the polarization tensor for GWB and 
${\bf D}_I$ is the detector tensor for each output signal, to which  
we specifically adopt the optimal TDI variables $A$, $E$ and $T$. 
The advantage of using the optimal TDIs in the cross-correlation analysis 
is that the noise spectra for cross-correlation data becomes exactly 
vanishing \cite{Prince:2002hp,Nayak:2002ir}. 
These variables are simply realized by 
combining the three Sagnac signals called S$_1$, S$_2$ and S$_3$ (often 
quoted as $\alpha$, $\beta$ and $\gamma$ in the literature): 
\begin{eqnarray}
&& {\bf D}_{\rm A} =  
\frac{1}{\sqrt{2} }( {\bf D}_{\rm S_3}- {\bf D}_{\rm S_1} ),
\nonumber
\\
&& {\bf D}_{\rm E} = 
\frac{1}{\sqrt{6}}(  
  {\bf D}_{\rm S_1} - 2{\bf D}_{ \rm S_2 } + {\bf D}_{ \rm S_3} ),
\nonumber
\\
&& {\bf D}_{\rm T} = 
\frac{1}{\sqrt{3}}( {\bf D}_{\rm S_1} 
+  {\bf D}_{\rm S_2} + {\bf D}_{\rm S_3} ).  
\label{eq:def AET mode}
\end{eqnarray}
The detector tensor for Sagnac signals can be obtained from the 
time-delayed combinations of one-way Doppler tracking calculations 
for optical-path length \cite{Cornish:2002rt,Cornish:2001bb,Kudoh:2004he}. 
For example, the Sagnac signal S$_1$ measures the phase difference 
between two laser beams received at space craft $1$, 
each of which travels around the LISA array in clockwise or 
counter-clockwise direction. Then, in the equal-arm length limit, 
the detector tensor for S$_1$ becomes  
\begin{eqnarray}
{\bf D}_{\rm S_1}({\bf\Omega},f;\,t) 
&=&  \frac{1}{2}\left[ \bigl\{{\bf a(t)}\otimes{\bf a(t)}\bigr\}\,
   {\cal T}_{\rm a}(\mathbf{\Omega},f;\,t) +
      \bigl\{{\bf b(t)}\otimes{\bf b(t)}\bigr\}\,
   {\cal T}_{\rm b}(\mathbf{\Omega},f;\,t) + 
    \bigl\{{\bf c(t)}\otimes{\bf c(t)}\bigr\}\,
   {\cal T}_{\rm c}(\mathbf{\Omega},f;\,t)
\right]~~;
\cr
{\cal T}_{\rm a}(\mathbf{\Omega},f;\,t) &=&
e^{-3i \hf/2 } 
\left\{
e^{  -  i (\hf/2) \, \{ -2 + {\bf a(t)}\cdot{\bf\Omega}  \}  }
  \sinc \left[ \frac{ \hf}{2}\left(  1 + {\bf a(t)}\cdot{\bf\Omega} 
\right) \right] 
- e^{- i (\hf/2) \, \{ 2 + {\bf a(t)}\cdot{\bf\Omega}\}  } 
  \sinc \left[ \frac{\hf}{2}  \left( 1-{\bf a(t)}\cdot{\bf\Omega} \right) \right]
\right\},
\cr
{\cal T}_{\rm b}(\mathbf{\Omega},f;\,t)
&=& 
 e^{- i (\hf/2) [3+\{{\bf a(t)}-{\bf c(t)}\} \cdot{\bf\Omega}]}
\left\{ 
    \sinc
    \left[ \frac{ \hf}{2 }\left(1 + {\bf b(t)}\cdot{\bf\Omega}\right) \right]  
  - \sinc \left[ \frac{\hf}{2 }\left(1-{\bf b(t)}\cdot{\bf\Omega} \right)
               \right] 
\right\},
\cr
{\cal T}_{\rm c}(\mathbf{\Omega},f;\,t) 
&=& 
e^{- 3 i\hf/2 } 
\left\{
e^{-  i (\hf/2) \{2 - {\bf c(t)} \cdot {\bf\Omega}\}} 
  \sinc \left[ \frac{\hf}{2}  \left(1+{\bf c(t)}\cdot{\bf\Omega}\right) \right] 
- 
e^{ i (\hf/2) \{2+{\bf c(t)}\cdot{\bf\Omega}\}}
  \sinc \left[\frac{\hf}{2}\left( 1-{\bf c(t)}\cdot{\bf\Omega}\right) \right]
\right\}, \nonumber
\end{eqnarray}
where the time-dependent vectors ${\bf a(t)}$, ${\bf b(t)}$ and ${\bf c(t)}$ 
represent the unit vectors pointing from the space craft $1$ to 
$2$, $2$ to $3$ and $3$ to $1$, respectively. Here, the quantity $\hf$ 
denotes the normalized frequency defined by 
\begin{eqnarray}
    \hf \equiv\frac{f}{f_*};\quad f_*=1/(2\pi\, L).
\end{eqnarray}
With the arm-length of LISA $L=5\times10^6$km, 
the characteristic frequency $f_*$ becomes $9.52$mHz. 
The analytic expressions for other detector 
tensors ${\bf D}_{\rm S_2}$ and ${\bf D}_{\rm S_3}$ are also obtained 
by the cyclic permutation of the unit vectors  
${\bf a}$, ${\bf b}$ and ${\bf c}$. 

In the expression of antenna pattern function (\ref{eq:def_antenna}) 
and/or detector tensor, 
the time-dependence is incorporated through the unit vectors 
${\bf a(t)}$, ${\bf b(t)}$ and ${\bf c(t)}$, which 
are determined by the orbital 
motion of LISA. Assuming the rigid motion in the equal-arm length limit, 
these are given by\footnote{Since we are interested in the directional 
information of gravitational-eve signals,  only the time evolution of 
directional dependence is considered. Note that the 
inclusion of radial dependence 
induces the Doppler modulation of the gravitational-wave signals, 
which is crucial for the identification of point sources. }
\begin{eqnarray}
\left(
\begin{array}{c}
{\bf a}(t) \\
{\bf b}(t) \\
{\bf c}(t) \\
\end{array}
\right)=\,{\bf R}_z(-\phi_D(t))\cdot{\bf R}_y(-\theta_D)\cdot
{\bf R}_z(\phi_D(t))\cdot
\left(
\begin{array}{c}
{\bf a}_0 \\
{\bf b}_0 \\
{\bf c}_0 \\
\end{array}
\right).
\end{eqnarray}
Here the quantities ${\bf R}_y$ and ${\bf R}_z$ are 
respectively  the $3\times3$ rotation matrices around $y$- and $z$-axes 
defined in the ecliptic coordinate system (see Fig.1 of paper I). 
The angles $\phi_D$ and $\theta_D$ are chosen as 
$\phi_D=-\pi+\omega_{\rm orbit}\,t$ and $\theta_D=-\pi/3$. The 
vectors ${\bf a}_0$, ${\bf b}_0$ and ${\bf c}_0$ represent the 
orientation of LISA in detector's rest frame. Here, we specifically set 
\begin{equation}
{\bf a}_0= \left(-\frac{\sqrt{3}}{2},~\frac{1}{2},~0\right),\quad
{\bf b}_0= \left(0,~-1,~0\right),\quad
{\bf c}_0= \left(\frac{\sqrt{3}}{2},~\frac{1}{2},~0\right).
\end{equation}

\begin{figure}[t]
\begin{center}
\includegraphics[height=2.5cm,clip,angle=0]{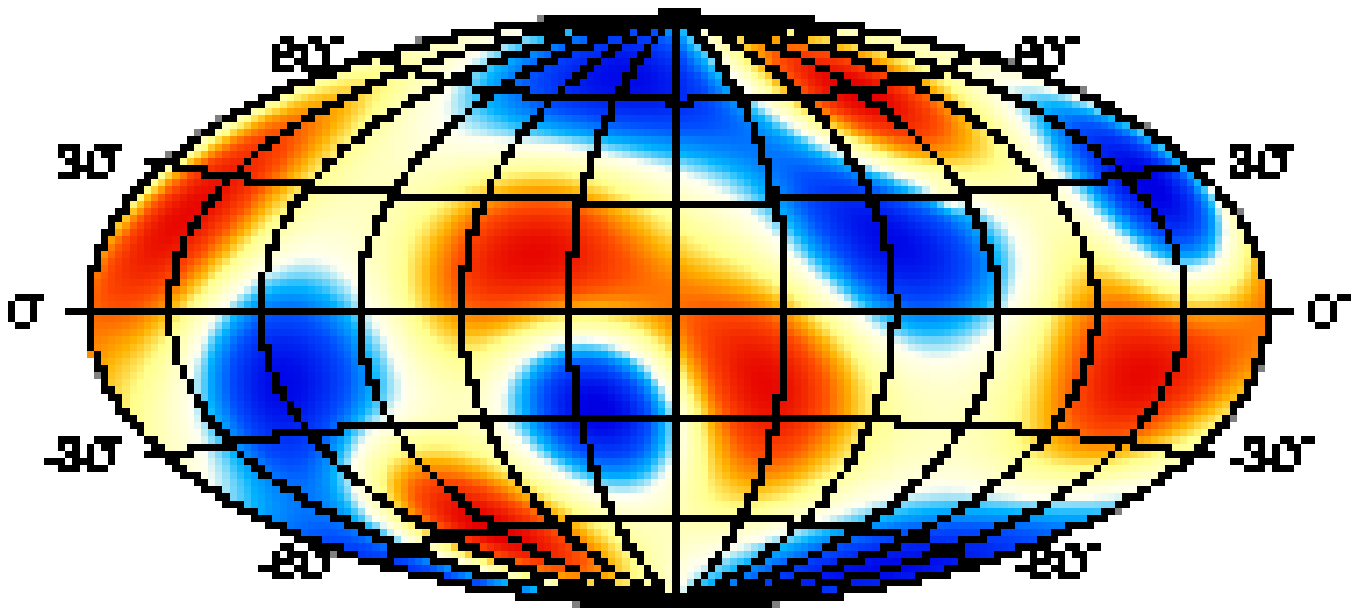}
\hspace*{0.5cm}
\includegraphics[height=2.5cm,clip,angle=0]{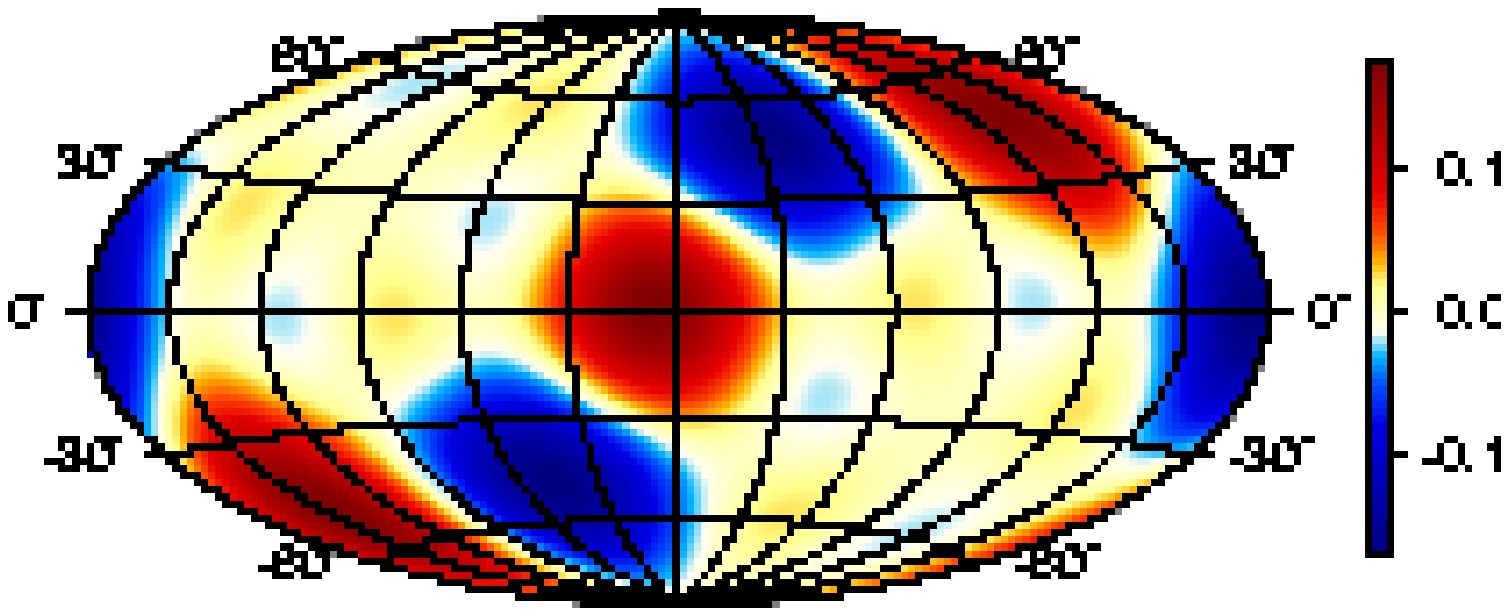}

\vspace*{0.6cm}

\includegraphics[height=2.5cm,clip,angle=0]{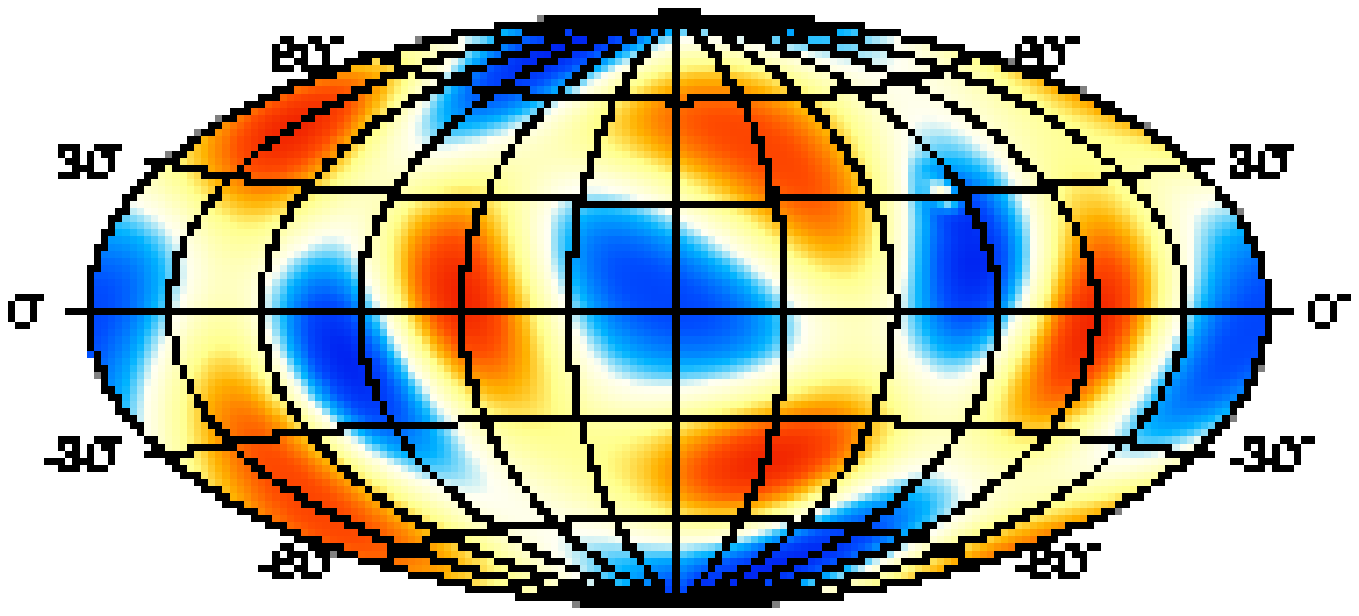}
\hspace*{0.5cm}
\includegraphics[height=2.5cm,clip,angle=0]{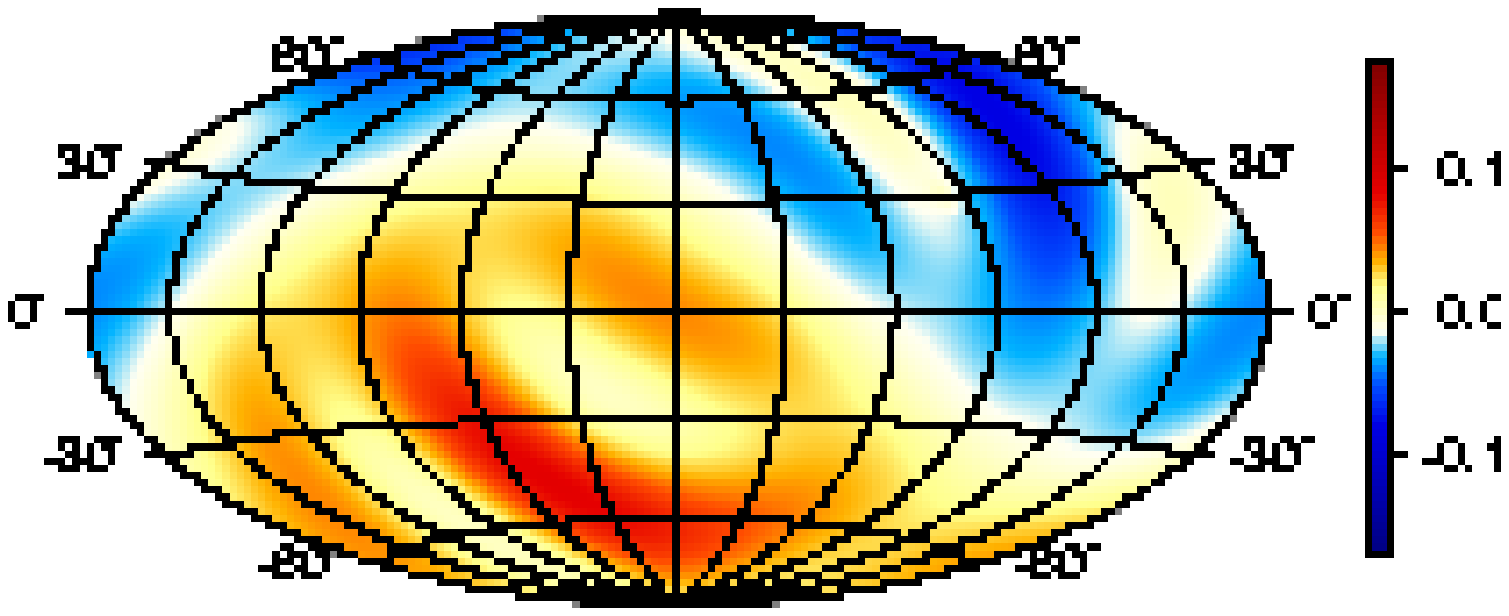}

\vspace*{0.6cm}

\includegraphics[height=2.5cm,clip,angle=0]{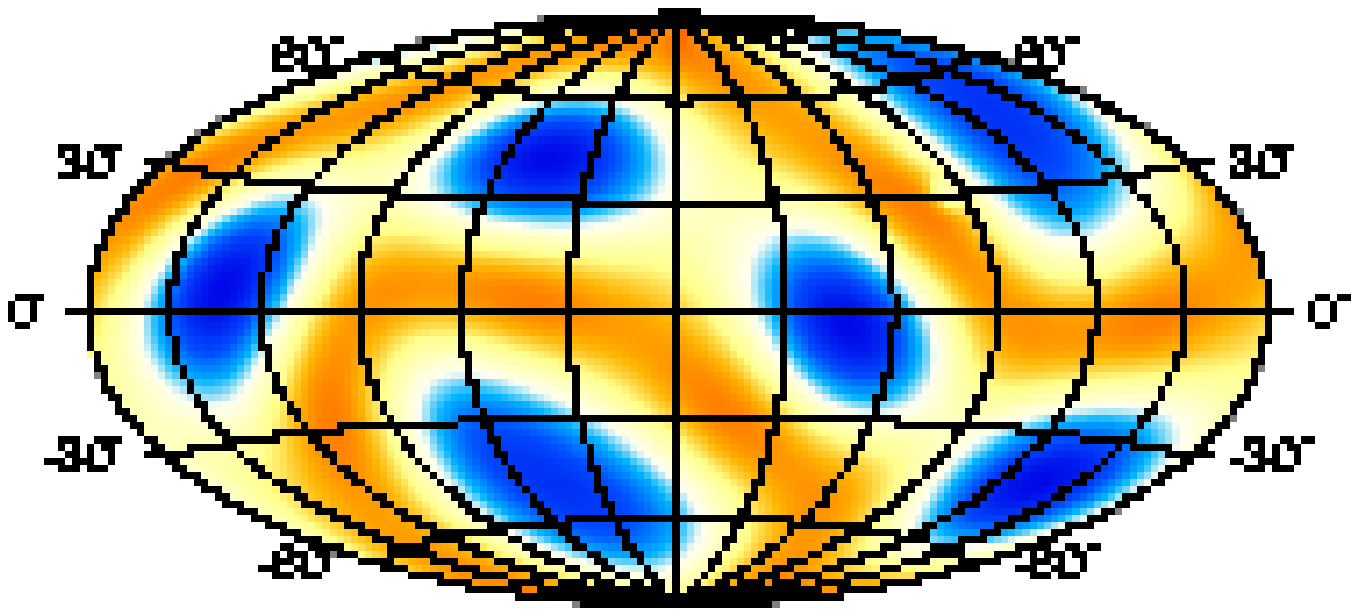}
\hspace*{0.5cm}
\includegraphics[height=2.5cm,clip,angle=0]{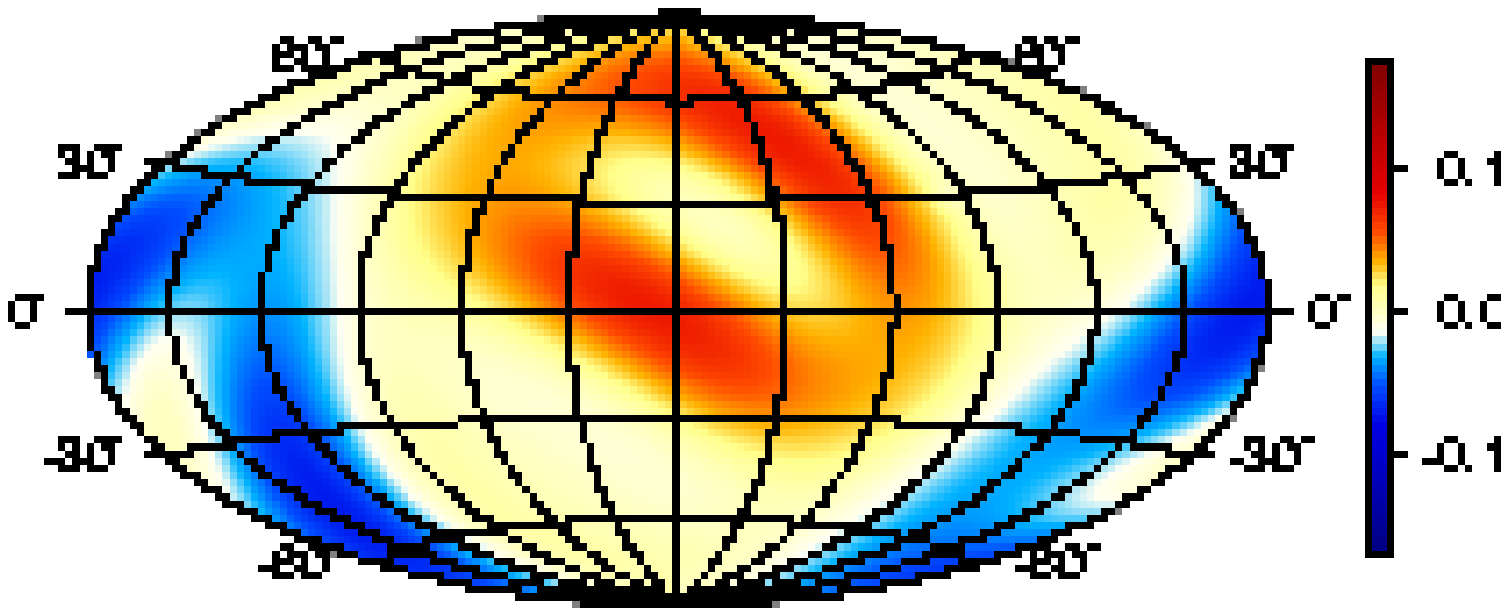}
\end{center}
    \caption{Snapshots of antenna pattern functions for 
      the cross-correlation signals 
      $AE$, $AT$ and $ET$ at the frequency 
      $f=3\,f_*\simeq28.6$mHz (from top to bottom). The left and the right 
      panels respectively show the real and the imaginary parts of 
      the antenna pattern functions. }
    \label{fig:antenna_pattern}
\end{figure}

Fig.~\ref{fig:antenna_pattern} shows the intensity distribution of 
the antenna pattern functions at $t=0$. The frequency is specifically 
chosen as $f=3f_*\simeq28.6$mHz. 
The antenna pattern functions at high-frequency regime give a 
different directional response, which implies that the cross-correlation 
signals $AE$, $AT$ and $ET$ mutually provide an independent 
information about source distribution. Note that the typical 
angular size of the intensity 
patterns is roughly $30^{\circ} \sim 60^{\circ}$, which 
basically limits the angular resolution of the final reconstructed skymap.

Finally, the noise spectral density for each output signal is presented, 
which is necessary to estimate the signal-to-noise ratios.  
Although the optimal TDI variables 
are free from the noise correlations, non-vanishing contribution to 
the self-correlation signals does exist.  
The noise spectral densities for optimal 
TDIs are calculated as (see also \cite{Cornish:2001bb,Prince:2002hp}):
\begin{eqnarray}
N_A(f) &=& N_E(f) 
= \sin^2(\hat{f}/2)\,\,\left\{8\,\left(2+ \cos\hat{f}\right)\,\,
S_{\rm shot}(f) + 16\,\left(3 +2 \cos\hat{f}+\cos2\hat{f} \right)\,\,
S_{\rm accel}(f)\right\},
\cr
N_T(f)&=& 2\,\,
\left(1+2\cos\hat{f}\right)^2\left\{ S_{\rm shot}(f) + 
4\sin^2(\hat{f}/2)\,\,
S_{\rm accel}(f) \right\},  
\label{eq:noise_TT}
\end{eqnarray}
where the variables $S_{\rm shot}$ and $S_{\rm accel}$ represent the 
proof mass noise and the optical path noise, respectively. 
Adopting the numerical values reported in 
Ref.\cite{Bender:1998} (see also \cite{Cornish:2001bb}), 
we obtain $S_{\rm shot}(f)=1.60\times10^{-41}$Hz$^{-1}$ and 
$S_{\rm accel}(f)=2.31\times 10^{-41}(\mbox{mHz}/f)^4$Hz$^{-1}$.




\end{document}